\documentclass[journal]{IEEEtran}
\usepackage{amsmath,graphicx,cite,epsfig}
\usepackage{multirow}
\usepackage{booktabs}
\usepackage{float}
\usepackage{xspace}
\usepackage{amssymb}
\usepackage{amsmath}
\usepackage{amsthm}
\usepackage{subfigure}
\usepackage{caption}
\usepackage{threeparttable}
\usepackage{diagbox}
\usepackage{url}
\usepackage{bm}
\usepackage{bm}
\usepackage{amsfonts}
\newcommand{\ieno}{\textit{i.e.}}
\newcommand{\egno}{\textit{e.g.}}
\newcommand{\tcr}{\textcolor{black}}
\newcommand{\ours}{FreqAlign}
\newcommand{\tcb}{\textcolor{black}}
\usepackage[table,xcdraw]{xcolor}
\usepackage{algorithm}  
\usepackage{algorithmicx}  
\usepackage{algpseudocode}
\usepackage{amsmath}
\usepackage{bbding}
\newcommand{\tct}{\textcolor{black}}
\ifCLASSINFOpdf

\else

\fi

\begin{document}
%
%
\title{FreqAlign: Excavating Perception-oriented Transferability for Blind Image Quality Assessment from A Frequency Perspective}
%
%

\author{Xin Li,~\IEEEmembership{Student Member, IEEE}, Yiting Lu, Zhibo Chen,~\IEEEmembership{Senior~Member,~IEEE}
\thanks{This work was supported in part by NSFC under Grant U1908209, 62371434, 62021001. (Xin Li and Yiting Lu contributed equally to this work.) (Corresponding author: Zhibo Chen.)}
\thanks{X. Li, Y. Lu, and Z. Chen are with the CAS Key Laboratory of Technology in Geo-Spatial Information Processing and Application System, University of Science and Technology of China, Hefei 230027, China (e-mail: lixin666@mail.ustc.edu.cn; luyt31415@mail.ustc.edu.cn; chenzhibo@ustc.edu.cn).}}  
\maketitle
\definecolor{mygray}{gray}{.9}
\begin{abstract}
Blind Image Quality Assessment (BIQA) is susceptible to poor transferability when the distribution shift occurs, \egno, from synthesis degradation to authentic degradation. To mitigate this, some studies have attempted to design unsupervised domain adaptation (UDA) based schemes for BIQA, which intends to eliminate the domain shift through adversarial-based feature alignment. However, the feature alignment is usually taken at the low-frequency space of features since the global average pooling operation. This ignores the transferable perception knowledge in other frequency components and causes the sub-optimal solution for the UDA of BIQA. To overcome this, 
from a novel frequency perspective, we propose an effective alignment strategy, \ieno, Frequency Alignment (dubbed \ours), to excavate the perception-oriented transferability of BIQA in the frequency space.
 Concretely, we study what frequency components of features are more proper for perception-oriented alignment. Based on this, we propose to improve the perception-oriented transferability of BIQA by performing feature frequency decomposition and selecting the frequency components that contained the most transferable perception knowledge for alignment. To achieve a stable and effective frequency selection, we further propose the frequency movement with a sliding window to find the optimal frequencies for alignment, which is composed of three strategies, \ieno, warm up with pre-training, frequency movement-based selection, and perturbation-based finetuning. 
Extensive experiments under different domain adaptation settings of BIQA have validated the effectiveness of our proposed method. The code will be released at \url{https://github.com/lixinustc/Openworld-IQA}.
\end{abstract}

.

\IEEEpeerreviewmaketitle

\begin{figure}
    \centering
    \includegraphics[width=0.95\linewidth]{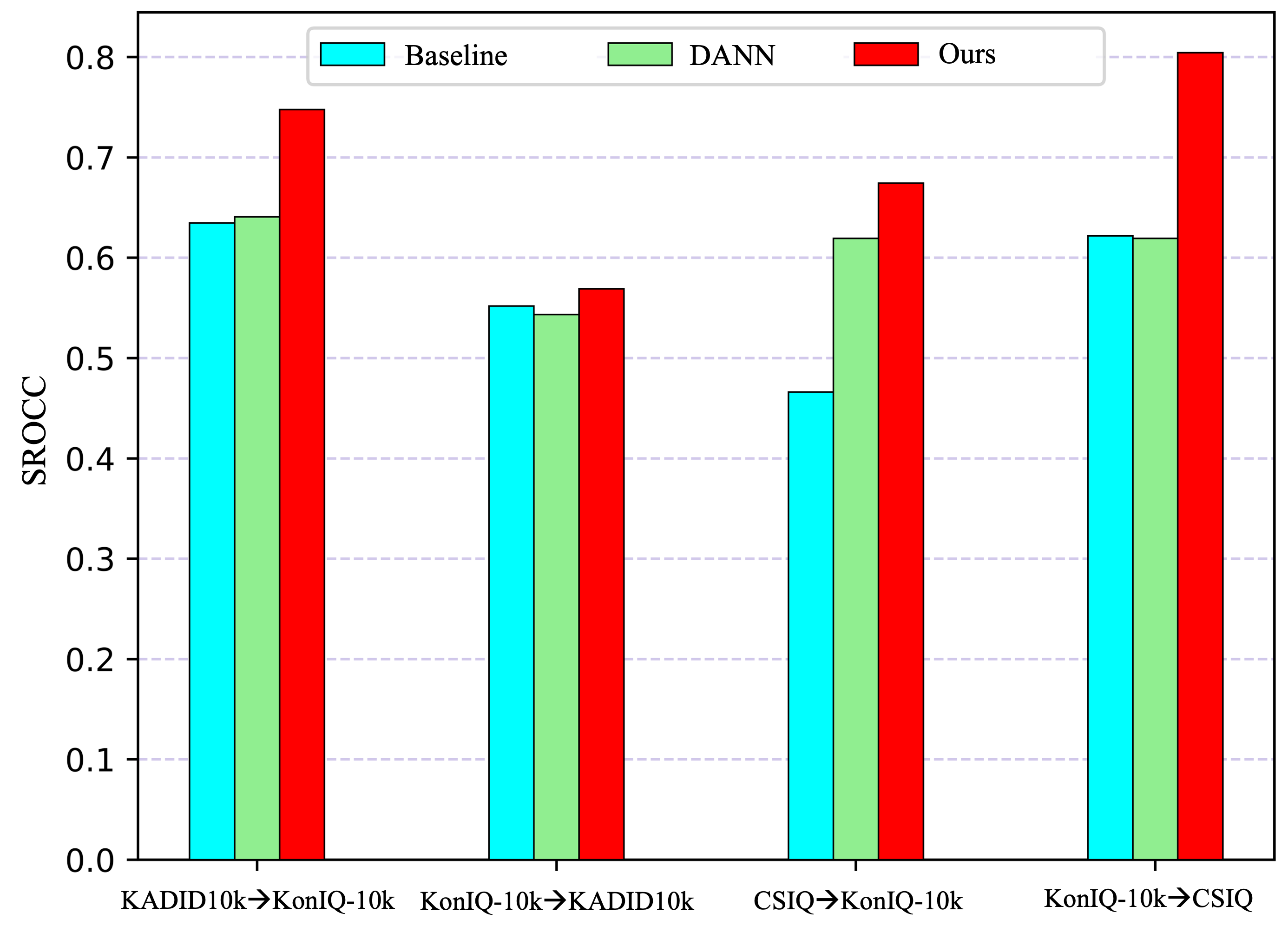}
    \caption{A comparison between our proposed FreqAlign, Baseline, and DANN~\cite{DANN} on target domain under four cross-domain settings, including two ``synthetic distortion source domain to one authentic distortion target domain", \ieno, KADID10k$\xrightarrow[]{}$KonIQ-10k and  CSIQ$\xrightarrow[]{}$KonIQ-10k, and two ``authentic distortion source domain to synthetic distortion target domain", \ieno, KonIQ-10k$\xrightarrow[]{}$KADID10k and KonIQ-10k$\xrightarrow[]{}$CSIQ. Here, ``Baseline" denotes that the BIQA model is directly trained with the labeled source domain data. ``DANN" is one general unsupervised domain adaptation method, which trains the BIQA model with labeled source domain data and unlabeled target domain data based on the adversarial-based feature alignment.}
    \label{fig:intro}
\end{figure}

\section{Introduction}
\label{sec:introduction}
\IEEEPARstart{R}{cent} years have witnessed the rapid progress of blind image quality assessment (BIQA) on general-purposed scenarios~\cite{zhai2020perceptual,yan2019naturalness,DBCNN}, and task-specific scenarios~\cite{JPEGcompressionIQA,JPEG2000IQA,blurIQA,gu2020giqa,jiang2022underwater}, \egno, generative task~\cite{gu2020giqa}, and compression~\cite{JPEGcompressionIQA,JPEG2000IQA}. 
\tct{However, an inevitable question is that a well-trained BIQA model might suffer from poor performance on the target domain (\ieno, the testing data) when the target domain has a large distribution gap with the source domain (\ieno, the labeled training data). As shown in Fig.~\ref{fig:intro}, ``Baseline" denotes the BIQA model is directly trained with the source data and validated on the target domain under severe distribution shift.  KADID10k$\xrightarrow[]{}$KonIQ-10k denotes the source domain is the synthetic distortion dataset KADID10k, and the target domain is the authentic distortion dataset KonIQ-10k. The well-trained BIQA model on the source domain data KADID10k only achieves about the SROCC of 0.6346 on the target authentic dataset KonIQ-10k due to its limited transferability.}
A na\"ive solution is to make annotations for target data, which is time-consuming and labor-intensive. To address this, unsupervised domain adaptation (UDA) has been introduced and delicately designed for BIQA, where the perception knowledge learned from labeled source data is expected to be transferred to the unlabeled target data by domain alignment.

There is a spectrum of works~\cite{UCDA,RankDA,SUFDA,lu2022styleam} exploring the application of UDA in BIQA to boost their transferability to new scenarios. The commonly-used schemes for UDA of BIQA can be divided into two categories, \ieno, discrepancy-based and adversarial-based. For instance, Chen \textit{et. al}~\cite{RankDA} minimize the MMD distance of ranked features from two distorted images to match the distribution between natural data and screen content data. UCDA~\cite{UCDA} conducts two-step adversarial alignment in curriculum style through uncertainty division, which achieves great transferability for video image quality assessment. Depart from that, to mitigate the domain gap between synthesized BIQA and authentic BIQA, Liu \textit{et al}.~\cite{SUFDA} is the first to propose the source-free UDA with a delicately designed self-supervised loss function-based on the intrinsic characteristic of BIQA. Lu \textit{et al}~\cite{lu2022styleam} aims to achieve the perception-oriented UDA by aligning the source and target domain in the style space. However, following the previous high-level UDA works, amounts of UDA works of BIQA still align the source and target domains at a coarse feature space, \ieno, the feature after average pooling operation, just like  Fig.~\ref{fig:frealign vs orialign} (a). Let us understand it from a novel frequency perspective~\cite{fcanet}. \textit{Average pooling operation is equivalent to extracting the low-frequency components of features}. And thus, previous works only take into account the transferable perception knowledge in the low-frequency component, which prevents adequate perception knowledge mining and transferring. 

\tct{In this paper, we aim to excavate the perception-oriented transferability for BIQA from a novel frequency perspective. Frequency space, as a  powerful representation space, has been widely exploited in a series of tasks, including IQA~\cite{marichal1999blur,muijs2005no,BLINDS,BLIINDS-II,DIIVINE}, classification~\cite{rao2021global,F_principle,highfrequencyhelps}, object detection~\cite{zhang2021deep}, pre-training~\cite{xie2022masked}, and image restoration~\cite{xie2021learning,pang2020fan,zou2022joint}. Among them, most works perform frequency decomposition at image space~\cite{xie2022masked,highfrequencyhelps} or feature space~\cite{xie2021learning,rao2021global,pang2020fan}. Existing frequency decomposition-based IQA methods~\cite{marichal1999blur,muijs2005no,BLINDS,BLIINDS-II,DIIVINE} are inspired by the observation that different frequencies are sensitive to different degradations, which motivated them to exploit the frequency components at image space for quality assessment. However, the above works are not designed to improve the transferability of the BIQA model and only exploit the correlation between different frequencies with different distortions. The investigation for the correlation between the transferability of BIQA and different frequency components at the feature level is significant and still under-explored.}

To explore the transferable perception knowledge that existed in different frequency components, we transform the features for domain alignment to a frequency space with the commonly-used Discrete Cosine Transform (DCT). Then, we perform the network training on source domain for each frequency component and test the perception performance on target domain.
The experimental results are shown in Fig.~\ref{fig:priliminary}, which indicates that the transferable perception knowledge does not only exist in the low-frequency component but also in the other frequency components of features. This interesting and significant finding leads us to further excavate the perception-oriented transferability from the powerful frequency perspective. 

Based on the above observation, we propose a novel strategy to take full advantage of the perception knowledge in multiple frequency components, \ieno, frequency alignment, which is shown in Fig.~\ref{fig:frealign vs orialign} (b). It is noteworthy that most unsupervised domain adaptation methods aim to learn the transferable knowledge from source data to target data by feature alignment, which narrows the domain shift and learns domain-invariant features. Therefore, we can further mine the transferable perception knowledge by aligning the frequency components of features of the source and target domains.
A na\"ive strategy is to perform frequency alignment for all frequency space. However, not all frequency is favorable for transferable perception knowledge mining. To select the most stable and efficient frequency components for alignment, we propose the frequency movement, which moves frequency components with a sliding window and estimates the transferability of each frequency band in a sequential manner (\ieno, following the predefined trajectory). 
The whole process of frequency movement is composed of three steps, including 1) warm-up with pre-training; 2) frequency movement-based selection; and 3) perturbation-based finetuning. Particularly, warm-up with pre-training aims to provide a well-trained backbone and discriminator, which prevents the side effects of the sub-optimal model for frequency selection in the second step. Perturbation-based finetuning can further adjust the frequency components for alignment after the second step, thereby providing the tolerance for the sub-optimal frequency selection.  Based on the above strategies, the feature extractor is able to learn transferable perception knowledge stably and does not have a bias for specific frequency components.  To validate the effectiveness of our \ours, we conducted experiments on three UDA settings in BIQA, including synthesis datasets to authentic datasets, authentic datasets to synthetic datasets, and multiple types of distortions to a single type of distortion. The final performance and related ablation studies have shown the superiority of our method. The contributions of this paper are summarized as follows:

The contributions of this paper are summarized as follows:
\begin{itemize}
    \item In this paper, we revisit the UDA of BIQA from the novel frequency perspective and pinpoint that previous works only conduct domain alignment at low-frequency components, ignoring the transferable perception knowledge in other frequency counterparts. 
    \item To take full advantage of transferable perception knowledge that existed in omni-frequency space, we propose the frequency alignment (\ieno, FreqAlign), which selects the optimal frequencies with frequency movement under the guidance of transferability and aligns them in an adversarial manner.  
    \item We conduct experiments on several typical UDA settings in BIQA, including the synthesized scenario to the authentic scenario, the authentic scenario to the synthesized scenario, and multiple distortions to one distortion. Abundant experiments and well-designed ablation studies have shown the superiority of our proposed FreqAlign. 
\end{itemize}

\section{Related works}
\label{sec:related work}
\subsection{Blind Image Quality Assessment}
Blind image quality assessment (BIQA) has played an irreplaceable role in academic and industry fields~\cite{madhusudana2022image,sun2022graphiqa,LIQA,zhai2020perceptual,liu2019unsupervised,CNNIQA,DBCNN,metaIQA,zhang2022continual,yang2020ttl,BLIINDS-II}. Since the lack of a reference image, BIQA  is a more challenging but more practical task~\cite{liu2019unsupervised}. In general, there are two directions for BIQA that are mostly investigated, \textit{i.e.,} BIQA for specific distortion~\cite{blurIQA,JPEG2000IQA,JPEGcompressionIQA} and general-purpose BIQA~\cite{NIQE,BLINDS,BLIINDS-II,BRISQUE,CNNIQA,DBCNN,yin2022CVR}. Typically, general-purpose BIQA aims to improve the robustness of the BIQA model for various distortions. Early works on BIQA usually adopt the handcraft perception features, \textit{e.g.,} scene statistics (NSS) features, based on  the assumption~\cite{NSS,ruderman1994statistics,srivastava2003advances} that the statistics extracted from the image will be changed by distortions. In recent years, deep learning has posed great progress in the BIQA~\cite{CNNIQA}, where the perception features are learned adaptively toward human perception from human-annotated datasets. The pioneering work CNNIQA~\cite{CNNIQA}, for the first time, introduces the convolution neural network to BIQA based on the synthetic distortion datasets~\cite{LIVE,ponomarenko2009tid2008}. However, since the domain gap between synthetic distortion and real-world distortion, the BIQA method performs poorly in practical applications. To solve the problem, DBCNN~\cite{DBCNN} adopts the two-branch architecture, where one branch is for the synthetic distortion, and another is optimized for the authentic distortion dataset based on the pre-training backbone with ImageNet. HyperIQA~\cite{HyperIQA} introduces the content understanding hyper network to boost the adaptability of the BIQA model for the distortions in the wild. \tct{Considering the varying size of images in the real world, MUSIQ~\cite{MUSIQ} proposes to utilize images at multiple scales and employ the hash-based 2D spatial embedding along with a scale embedding to better perceive position and scale information from multi-scale inputs.
To investigate the relationship between multiple distortions and the perceptual quality, GraphIQA \cite{sun2022graphiqa} constructs a distortion graph to extract the distributional characteristics of different distortion types and levels, thereby improving the regression between distortions and quality score. 
In contrast, VCRNet~\cite{pan2022vcrnet} enables the BIQA model to better perceive the distortions by introducing the difference between the distorted images and their corresponding restored images. They introduce a non-adversarial restoration model to compensate for the destroyed low-level semantic features and high-frequency details.}

To unify the BIQA for the synthetic and authentic distortions, Zhang \textit{et al.}~\cite{zhang2021uncertaintyIQA} exploit the shared uncertainty-based rank knowledge existing in the synthetic and authentic distortions to achieve general BIQA. With the development of meta-learning, MetaIQA~\cite{metaIQA} utilizes it to learn a prior model that contains the meta-knowledge shared by the human perception process, which can be adapted to unseen distortions efficiently. \tct{Moreover, to increase the continual learning capability of the BIQA model for sequential BIQA tasks, and prevent catastrophic forgetting of historical BIQA tasks,
Zhang \textit{et al.}~\cite{zhang2022continual} proposes a pseudo-label-based regularization strategy to consolidate previous knowledge and a multi-head aggregation strategy to reduce task-dependent bias. }
 In this paper, we aim to achieve the UDA of BIQA, where the labels of target data are not accessible.

\begin{figure*}[htp]
    \centering
    \includegraphics[width=1\textwidth]{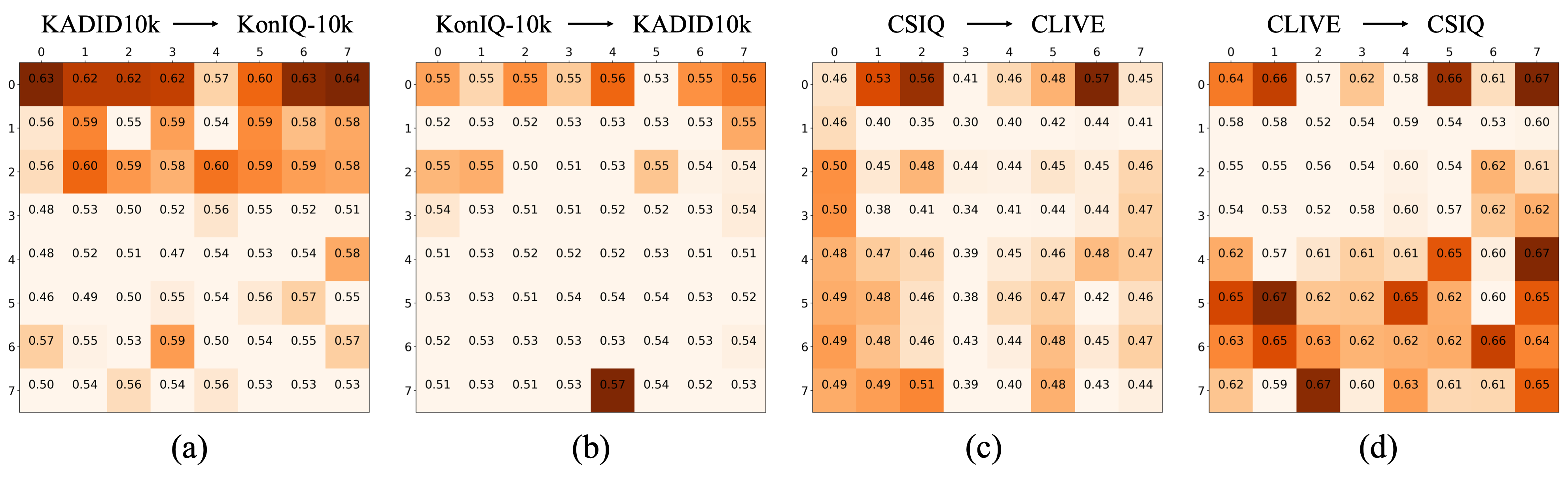}
    \caption{\tct{Visualization of the perception-oriented transferability in different frequency components. The experiments are conducted in four cross-domain settings, including two synthetic datasets (\ieno, KADID10k~\cite{Kadid10k} and CSIQ~\cite{CSIQ}) and two authentic datasets (\ieno, KonIQ-10k~\cite{KonIQ}  and CLIVE~\cite{LIVEC}.) Higher SROCC value represents higher perception-oriented transferability. The frequency matrix is obtained with the DCT transform. The lowest frequency component and highest frequency component are in the upper left corner and lower right corner of the grid, respectively.} }
    \label{fig:priliminary}
\end{figure*}

\subsection{Unsupervised Domain Adaptation}
  Unsupervised Domain Adaptation (UDA)~\cite{mmd,ADDA,CORAL,DANN1,CCD,cmmd,metaalign,jin2021re,yang2023tvt,yu2023classification,zhang2021self} aims to transfer knowledge from the source domain to the target domain with the help of labeled source data and unlabeled target data. The principle of UDA is to learn the domain invariant knowledge between the source and target domains by reducing their domain divergence~\cite{zhang2021survey}. The commonly-used methods for UDA can be roughly divided into two categories, including feature alignment-based and self-training-based. Among them, feature alignment is usually achieved by reducing the hand-crafted statistic distances of features~\cite{mmd,mmd1,cmmd,CORAL,Wa_distance}, \egno, MMD, or reducing the feature distance in an adversarial manner~\cite{DANN1,ADDA,metaalign}. There are two categories of methods for self-training-based UDA. The first category~\cite{yang2021st3d,li2022semantic,liu2021cycle,wang2022uncertainty} exploits the pre-trained backbone to produce the pseudo-labels for target data, and then finetune them with labeled target data.  The second one~\cite{yang2023tvt,SUFDA} designs the self-supervised loss for the target data, which is used to optimize the backbone together with the source data.

  Recently, some works have taken a step forward to investigate the UDA of quality assessment~\cite{chen2021no,UCDA,wang2021learningOpinion,SUFDA,lu2022styleam,tliba2022deepmedical,nabavi2021automaticmedical,yang2022no}, which aims to transfer the perception knowledge from the source domain to the target domain. Following the UDA on high-level tasks, Chen \textit{et al.}~\cite{chen2021no} leverage the maximum mean discrepancy to align the ranked features, thereby achieving the perception knowledge transfer between natural images and screen content images.  UCDA~\cite{UCDA} proposes to divide the target data into certain and uncertain subdomains, where the certain subdomain is aligned with self-supervised loss and another one is aligned with adversarial loss. Wang \textit{et al.}~\cite{wang2021learningOpinion} achieve the opinion-free BIQA by the unsupervised domain adaptation between paired source and target data. To further improve the transferability of BIQA, StyleAM~\cite{lu2022styleam} investigates the correlation between the feature style and quality score, and proposes the style alignment for the UDA of BIQA. SFUDA~\cite{SUFDA} achieves the source-free domain adaptation for BIQA by introducing the self-supervised losses based on the quality distribution. \tct{Compared with existing UDA-based IQA methods, which only exploits the transferable perception knowledge in low-frequency component, our FreqAlign first pinpoint that the transferable perception knowledge of BIQA is distributed in multiple frequency components of features, and we propose the frequency alignment based on sliding window-based frequency movement, adequately excavating the transferable perception knowledge existing in omni-frequency space.}

\subsection{Frequency Learning}
Frequency space has been explored to boost the deep learning in various tasks, such as image classification~\cite{rao2021global,F_principle,highfrequencyhelps}, visual pre-training~\cite{xie2022masked}, and image processing~\cite{xie2021learning,pang2020fan,li2021learning,zhang2023cfpnet}. The commonly-used strategies are to decompose the images into different frequencies and process/enhance them in a divide-and-conquer manner. Recently, one fascinating work~\cite{F_principle} finds that convolution neural networks (CNNs) tend to learn features from low-frequency to high-frequency for specific tasks. This prevents the generalizations of CNNs to the datasets where the high-frequency features matter. Following this, a series of studies~\cite{zhang2022spectral,huang2021rda,yang2020fda,Amplitude_spectrum_segmentation} have been proposed to improve the generalization capability of models for high-level tasks
from the frequency perspective, which can be roughly divided into two categories. The first category of works aims to decompose the features into different frequencies, in which the domain-invariant components are enhanced and domain-variant components are eliminated/perturbed with an attention mechanism~\cite{zhang2022spectral} or gating mechanism~\cite{huang2021rda}.   
 The second category of works~\cite{yang2020fda,Amplitude_spectrum_segmentation} is mostly based on the observation that the amplitude of frequencies is correlated with domain style, while the phase of frequencies contains more semantics that is domain invariant. Therefore, the domain augmentation strategy is proposed to swap the styles of two images by decomposing the images into the frequency space and swapping their amplitudes, which enforces the network to narrow the domain gap and improve the generalization capability. However, the above works on generalization mostly focus on semantic tasks, which is not proper for the UDA of BIQA. In contrast, we aim to excavate the perception-oriented transferability for BIQA through a frequency perspective. 

\tct{Frequency decomposition has been exploited in several IQA methods~\cite{marichal1999blur,muijs2005no,BLINDS,BLIINDS-II,DIIVINE}. These methods are based on the observation that different frequencies are sensitive to different distortions, which can be roughly divided into two prominent categories: 1)  The first category~\cite{marichal1999blur,muijs2005no} performs the quality assessment on the specific frequency component based on the intuition of the relation between distortions and frequency (\ieno, the blur will destroy the high frequency mostly) in a hand-crafted manner. For instance, Marichal \textit{et.al}~\cite{marichal1999blur} argues that the blur artifacts are correlated with high frequencies, thereby estimating blur degrees based on histograms of non-zero DCT components   2) 
The second line~\cite{BLINDS,BLIINDS-II,DIIVINE} utilizes full-frequency statistics of the original image for the regression of the quality score, which enables the general-purpose IQA model to be adaptive to multiple degradations. In particular, BLIINDS~\cite{BLINDS} and BLIINDS-II~\cite{BLIINDS-II} extract quality-related features from full-frequency statistics and then utilize a probabilistic model to map them to quality scores. However, the above works are only devoted to improving the quality assessment of seen domain data, which ignores the transferability of BIQA. In contrast, we aim to investigate the transferability of the perception knowledge of different frequencies, which is exploited to improve the transferability of existing BIQA methods for out-of-distribution (OOD) BIQA data.}

\begin{figure*}[htp]
    \centering
\includegraphics[width=0.95\textwidth]{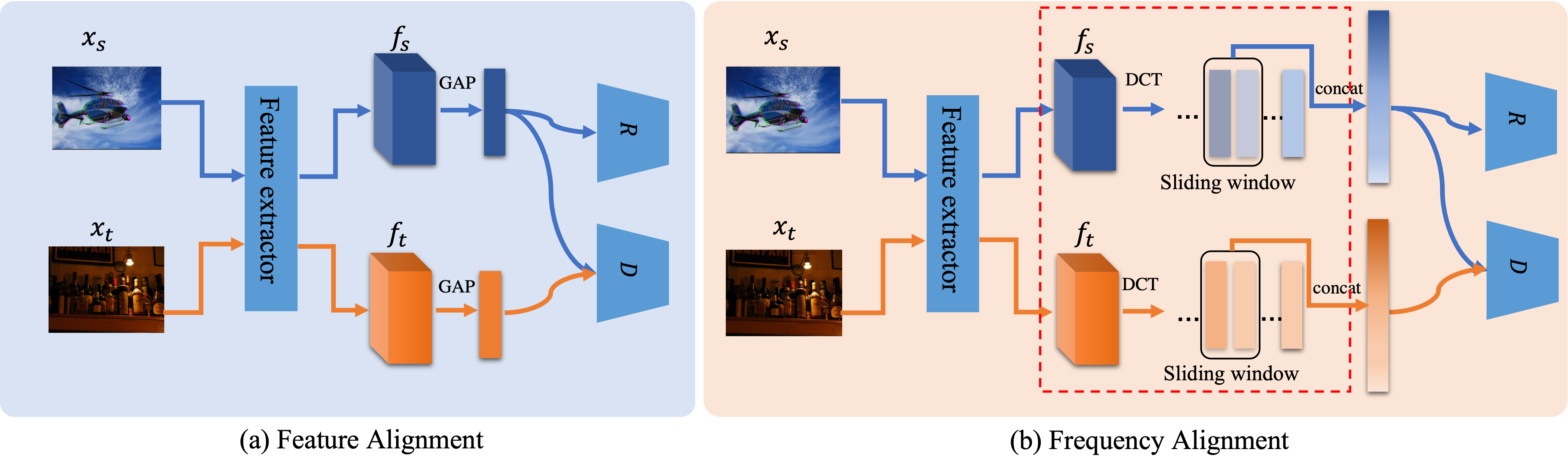}
    \caption{A comparison between (a) feature alignment and (b) our frequency alignment. Our frequency alignment is conducted on the multiple frequency components of the source feature $f_s$ and target feature $f_t$ in the sliding window to mine perception knowledge, while feature alignment is being used to align the low-frequency component of feature $f_s$ and $f_t$ due to the use of a GAP (global average pooling) operation. Here, $\mathrm{GAP}$ refers to global average pooling, $R$ and $D$ denotes the regression head and discriminator.}
    \label{fig:frealign vs orialign}
\end{figure*}
\section{Approach}
In this paper, we aim to excavate transferable perception knowledge for BIQA from the frequency perspective. In Sec.~\ref{sec:prob definition}, we first introduce the problem definition for the unsupervised domain adaptation of BIQA. Then we revisit the UDA of BIQA in frequency space in Sec.~\ref{B}, which reveals why perception knowledge transfer in frequency space is necessary. To sufficiently excavate the transferable perception knowledge, we propose the effective and efficient frequency movement-based alignment in Sec.~\ref{C}.

\subsection{Problem Definition}
\label{sec:prob definition}
Unsupervised Domain Adaptation (UDA) of BIQA aims to excavate the transferable perception knowledge from labeled source domain data $D_s=\{x_i^s, y_i^s\}_{i=1}^{n_s}$ for the target domain data without access to their labels $D_t=\{x_j^t\}_{j=1}^{n_t}$. Here, $x_i^s$ and $y_i^s$ denote the sample and its corresponding quality score of the source domain, respectively. $n_s$ and $n_t$ are the numbers of the samples of the source and target data. To eliminate domain shift, the commonly-used algorithm exploits the adversarial-based feature alignment~\cite{DANN1} to narrow the distances between two domains, which enables better perception knowledge transfer. Particularly, one discriminator $D$ is optimized to distinguish the source and target domains with a domain classification loss. Then the generator (\textit{i.e.,} feature extractor) $G$ is enforced to learn the domain-invariant perception knowledge, which is optimized by minimizing the quality aggregation loss $\mathcal{L}_S$ with regression head $R$ and maximizing the domain classification loss $\mathcal{L}_D$ with discriminator $D$:
\begin{equation}
   \theta_G = \arg\min_{G}\mathcal{L}_S-\mathcal{L}_D
\end{equation}
 The most crucial challenge of this task stems from ``how to identify and make full use of the transferable perception knowledge existed in the source domain"

\subsection{Revisiting the UDA of BIQA in Frequency Space}
\label{B}
As the pivotal technology in UDA, feature alignment is the core reason for excavating the transferable knowledge from source data to target data by eliminating the domain shift. Early works~\cite{mmd,CORAL,cmmd} usually achieve the feature alignment with the hand-craft distribution metric, \egno, MMD, which lacks enough flexibility and adaptability. To overcome this, adversarial-based feature alignment~\cite{DANN1,ADDA} is proposed to measure the distribution distance and narrow it adaptively. However, the above strategies excessively focus on distribution alignment, while ignoring task-related knowledge (\ieno, perception knowledge in BIQA).

Let us revisit the feature alignment in the UDA of BIQA. Following the UDA of classification tasks~\cite{DANN1,ADDA,metaalign}, the commonly-used feature alignment strategies of BIQA are achieved at the compact feature space after the Average Pooling Layer~\cite{lin2013network}, which is shown in Fig.~\ref{fig:frealign vs orialign}(a). However, from the frequency perspective, this operation only exploits the low-frequency component of features to excavate the perception knowledge~\cite{fcanet}. Different from the UDA of classification, where the semantics of each class are more likely to exist in the global features~\cite{shang2020multi,chen2020global,luo2022frequency} (\ieno, the low-frequency features), the perception knowledge of BIQA is distributed in multiple frequency components of features.

In this paper, we make a comprehensive experimental analysis for the transferable perception knowledge existing in different frequency components. To ensure the reliability of our experiments, we explore their dependency on two typical cross-domain scenarios in the UDA of BIQA: 1) synthetic distortion (\egno, KADID10k dataset~\cite{Kadid10k} and CSIQ dataset~\cite{CSIQ}) to authentic distortion (\egno, KnoIQ-10k dataset~\cite{KonIQ} and CLIVE dataset~\cite{LIVEC}) and 2) authentic distortion to synthetic distortion.
Concretely, we perform the Discrete Cosine Transform (DCT) to the features after the last layer on the spatial dimension, which results in the 8$\times$8 frequency matrix as Fig.~\ref{fig:priliminary}. The lowest frequency component is in the upper left corner of the grid. From left to right and top to bottom, the frequency component becomes higher in the horizontal and vertical dimensions.   
To measure the transferable perception knowledge in each frequency component, we adopt subjective quality regression on the source domain for each frequency component and test performance on the target domain. The perception results  ``SROCC” of all frequency components on the target domain are presented with the grid in Fig.~\ref{fig:priliminary}, where more bright color means a higher perception transferability. From Fig.~\ref{fig:priliminary}, we can obtain two crucial findings as: 1) \textbf{The most effective frequency component for excavating transferable perception knowledge is contingent upon different scenarios.} (\egno, the optimal frequency component is at (0, 7) for KADID10k $\xrightarrow[]{}$ KonIQ-10k but (7, 4) for KonIQ-10k $\xrightarrow[]{}$ KADID10k in Fig.~\ref{fig:priliminary}). 2) The frequency components with comparable performance tend to cluster around. The above findings encourage us to develop an effective feature alignment strategy (\textit{frequency movement-based alignment}) for the UDA of IQA.

\begin{figure*}
    \centering
\includegraphics[width= 0.92\textwidth]{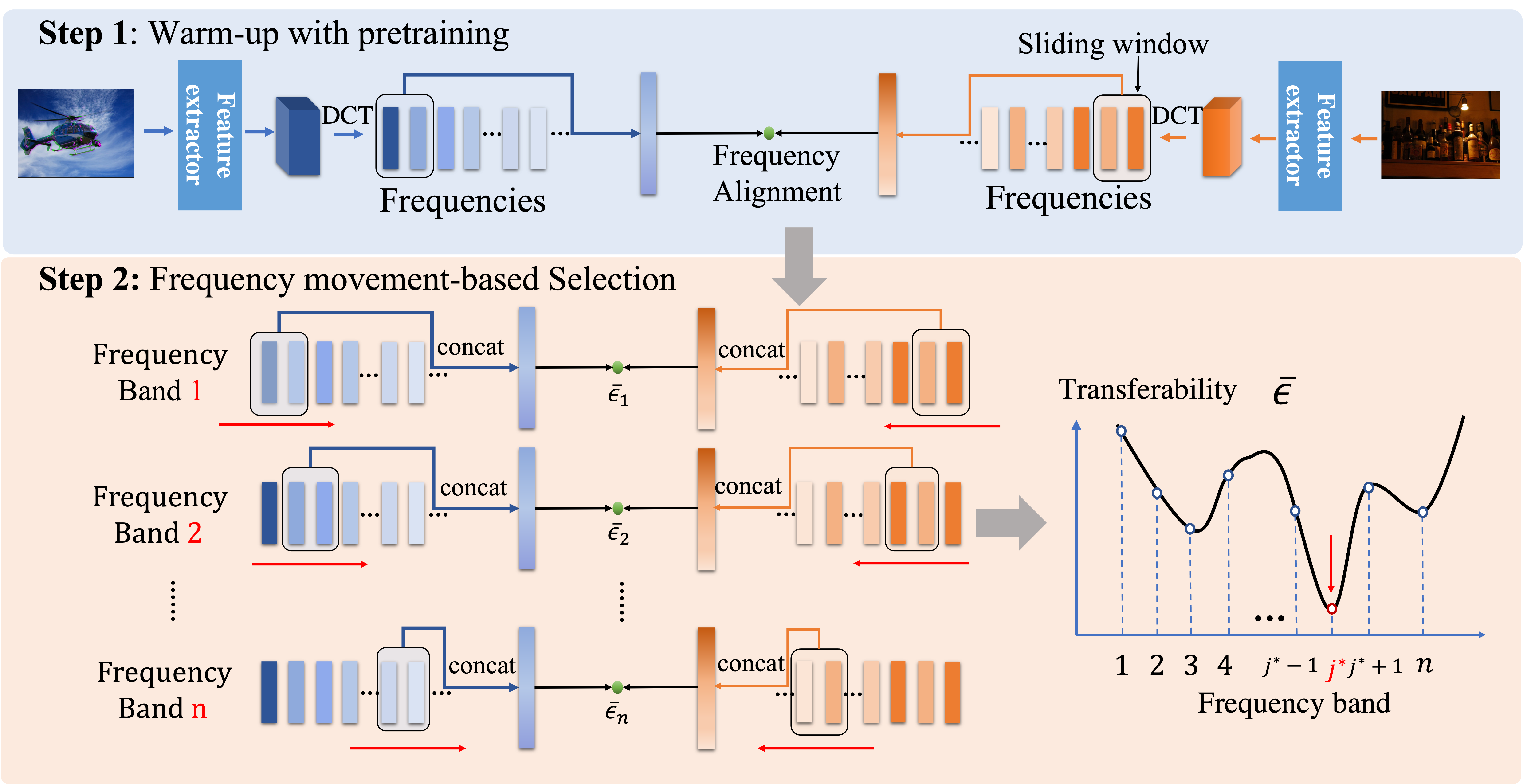}
    \caption{The overview of the first two steps in the proposed frequency movement-based alignment. In the first step, the warm-up with pre-training is achieved by the frequency alignment in the first frequency band. In the second step, a frequency movement-based selection is achieved by moving the frequency band in a window together with frequency alignment.}
    \label{fig:stylemovement}
\end{figure*}
\subsection{Frequency Movement-based Alignment}
\label{C}
To further excavate perception knowledge for unlabeled target data, we aim to achieve the UDA of BIQA from the frequency perspective. There is one crucial challenge for this purpose: \textit{How to select the optimal frequency components for alignment.} 
 One na\"ive solution is to conduct the alignment for each frequency component individually and select the optimal one. However, it will cause an unaffordable cost of time and resources since abundant frequency components. We aim to integrate optimal frequency selection and frequency alignment in the same training process. To achieve sufficient and stable frequency selection, we propose frequency movement with a sliding window, which is composed of three steps, and the first two steps are shown in Fig.~\ref{fig:stylemovement}. In the first step, we achieve the warm-up for the backbone and discriminator with the pre-training at zero frequency component. This ensures stable frequency selection in the process of frequency movement. In the second step, the frequency selection is achieved by jointly moving the frequencies with a sliding window forward following the predefined trajectory and obtaining the transferability with frequency alignment. After obtaining the initial optimal frequencies, we propose perturbation-based finetuning to achieve the final frequency alignment in the third step. In this way, we can excavate the more abundant transferable perception knowledge for BIQA. 
 More technique details are described as follows:

\subsubsection{Frequency Decomposition on Deep Feature Space}
It is noteworthy that the perception knowledge transfer is achieved in the feature space. And thus, we conduct the frequency decomposition for BIQA on the deep feature space. Specifically, given the source image $x_s$ and target image $x_t$, we first extract the deep feature $f_s=G(x_s)\in \mathbb{R}^{C\times H \times W}$ and $f_t=G(x_t)\in \mathbb{R}^{C\times H \times W}$from the last convolution layer of feature extractor $G$. Then we decompose the features into multiple frequency components through 2D discrete cosine transform (DCT) for each channel. Let us represent the bias function of 2D DCT as:
\begin{equation}
    B_{h,w}^{i, j} = cos(\frac{\pi i}{H}(h+\frac{1}{2}))cos(\frac{\pi j}{W}(w+\frac{1}{2}))
\end{equation} 
Then, the decomposed frequency components of features $f_s$, and $f_t$ can be calculated as:
\begin{equation}
    \begin{split}
        \mathcal{F}_s^{i, j} = \sum_{h=0}^{H-1}\sum_{w=0}^{W-1}f_s^{h, w}B_{h, w}^{i, j} \\ 
    \mathcal{F}_t^{i, j} = \sum_{h=0}^{H-1}\sum_{w=0}^{W-1}f_t^{h, w}B_{h, w}^{i, j},
    \end{split}
    \label{eq:dct_transform}
\end{equation}
where $\mathcal{F}_s^{i, j}$ and $\mathcal{F}_t^{i, j}$ denote elements in the $i^{th}$ row and the $j^{th}$ column of the decomposed frequency matrix of the source and target data, respectively. We remove the constant coefficients in the Eq.~\ref{eq:dct_transform} for a simple description.  With the increase of the $i$ and $j$, the frequency varies from low to high in the horizontal and vertical directions.

\subsubsection{Frequency Movement}
After decomposing the features into frequency space, one straightforward strategy is to align all frequency components of the source and target data. However, not all frequency components contain useful perception knowledge for transfer. Simply aligning all frequencies will cause a cost of time and resources and even lead to side effects. To achieve a stable and efficient frequency selection and alignment, we propose frequency movement, which moves the frequency band in a sliding window forward in the process of frequency alignment.  In this way, we can obtain the perception-oriented transferability of each frequency band with a shared feature extractor and discriminator, and then, select the optimal one.  However, this brings two crucial challenges: 1) How to find one optimal movement trajectory. 2) How to find one optimal metric to measure perception-oriented transferability.

\begin{figure*}
    \centering
\includegraphics[width=0.8\linewidth]{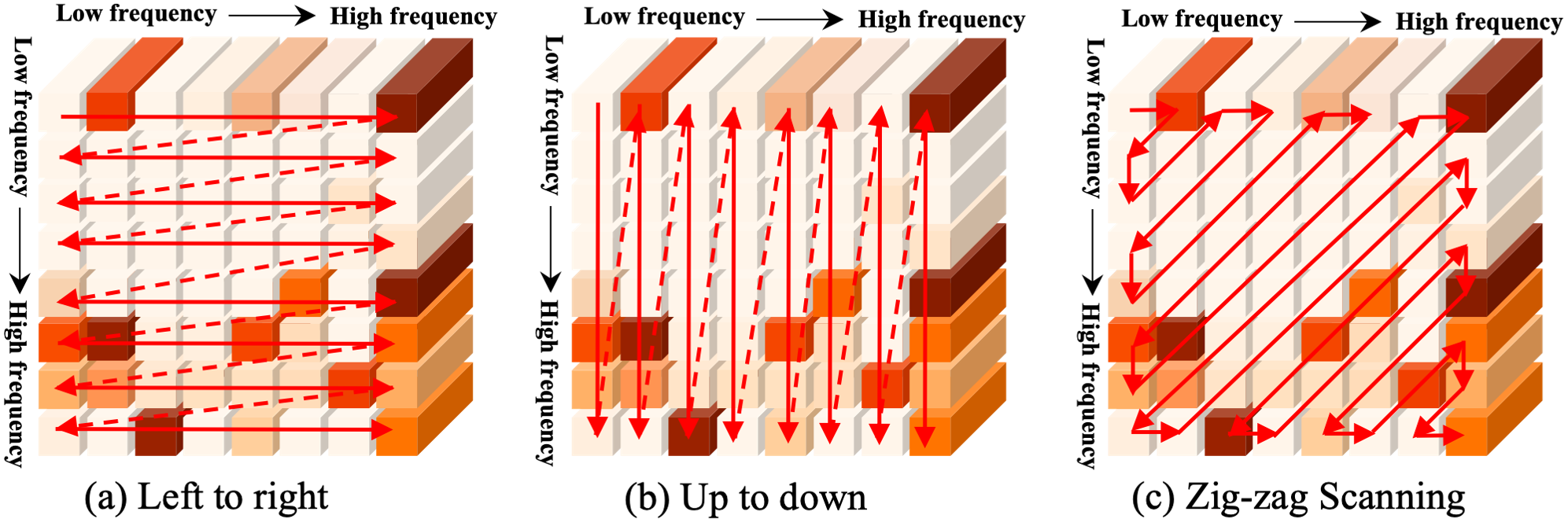}
    \caption{The visualization of different frequency movement trajectories, including (a) left-to-right, (b) up-to-down, and (c) Zig-zag scanning. }
    \label{fig:movement_trajectory}
\end{figure*}

For the first challenge, we investigate three typical trajectories in Fig.~\ref{fig:movement_trajectory} for frequency movement, \ieno,  1) moving frequency from left to right along the row, 2) moving frequency from up to down along the column, and 3) moving frequency with the Zig-zag Scanning. There is another crucial question is whether one frequency in the band is enough for perception knowledge transfer. Intuitively, the perception knowledge of different frequencies might be complementary. Therefore, we propose the multiple frequencies movement. As shown in Fig.~\ref{fig:stylemovement}, we set a sliding window with the size of $m$, which covers the $m$ continuous frequencies to form a frequency band. Then we utilize the frequency band in the sliding window for frequency alignment and quality regression.
The $j$-th frequency band of the source domain and target domain can be represented as $\widetilde{\mathcal{F}}_s^{j}$ and $\widetilde{\mathcal{F}}_t^{j}$:
\begin{equation}
   \begin{split}
    \widetilde{\mathcal{F}}_{s}^{j}&= [\mathcal{F}_{s}^j,\mathcal{F}_{s}^{j+1},...,  \mathcal{F}_{s}^{j+m-1}] \\
    \widetilde{\mathcal{F}}_{t}^{j}&= [\mathcal{F}_{t}^j,\mathcal{F}_{t}^{j+1},...,  \mathcal{F}_{t}^{j+m-1}] \\
    \end{split}
\end{equation}
With the increase of iterations, the sliding window will move forward following the predefined trajectory. 

For the second challenge, we are required to find one optimal metric to decide how to conduct selection during frequency movement. And the metric requires to have the capability to measure the transferability. There are three commonly-used metrics for the measurement of transferability, including MMD~\cite{mmd}, CORAL~\cite{CORAL}, and domain adversarial loss~\cite{DANN1}. We systematically investigate these three typical metrics and experimentally find all of these metrics can guide the frequency movement to obtain excellent performance, of which the MMD achieves slightly better performance. Therefore, we utilize the MMD as the metric for the transferability measurement.

Based on the above strategies, in this paper, we take three steps to achieve the frequency movement, \textit{i.e.,} \textit{warm-up with pre-training}, \textit{frequency movement-based selection}, and \textit{perturbation-based alignment}.

\noindent\textbf{\textit{Warm-up with pre-training.}}
\tcr{In general, inadequate training of the backbone will suffer in inaccurate frequency selection, where the potential early frequency bands are easily overlooked. Consequently, at the first step, a warm-up for the BIQA model and discriminator is required to obtain the well-trained initial model for frequency selection. The model is warmed up at the first frequency bands to let the transferability metric $\epsilon$ be stable, which will prevent the effects of the inadequate training of the backbone.}

\noindent\textbf{\textit{Frequency Movement-based Selection.}}
In the second step, we aim to select the optimal frequency band for frequency alignment with frequency movement.  \tct{As shown in Fig.~\ref{fig:stylemovement}, following the predefined trajectory, we move the frequency band with the sliding window and align the source and target domains at this frequency band to measure the transferability of this frequency band $\overline{\epsilon}$. However, the measurement of transferability is susceptible to the fluctuation caused by the randomness in selected samples and network optimization at each step. It is not accurate to utilize the MMD distance at one iteration step as the measurement of transferability. To make it reliable, we average the measured MMD distance for the $j^{th}$ frequency bands in $T$ iterations, which avoids the fluctuation for the measurement of transferability caused by the randomness in sample selection and network optimization. Therefore, we can measure the transferability of the $j^{th}$ frequency band as:
\begin{equation}
\overline{\epsilon}_j=\frac{\sum_{t=j\times T}^{(j+1)\times T}\epsilon(t)}{T},
\label{eq:epsilon}
\end{equation}
where $\epsilon(t)$ denotes the MMD distance between the features of source data and target data at $t$ timestamp, and $T$ is the time interval (\ieno, iterations) for each measurement.}

With the above strategy, we can measure the transferability $\overline{\epsilon}$ of all frequency bands by traversing overall frequency bands with the sliding window as:
\begin{equation}
\overline{\epsilon}=[\overline{\epsilon}_1,\overline{\epsilon}_2,..,\overline{\epsilon}_j, ...]
\end{equation}
The order of the optimal frequency band is obtained by $j^{*}=argmax(\overline{\epsilon})$.

\noindent\textbf{\textit{Perturbation-based Finetuning.}}
 After obtaining the optimal frequency band, one intuitive idea should be finetuning the whole BIQA model with this frequency band. However, it lacks the tolerance capability for unstable frequency selection in the second step. To avoid this, we introduce frequency perturbation-based finetuning for frequency movement. Specifically, with the ${j^*}^{th}$ selected frequency band as the center, we perturb the frequency band within an interval $[j^*-k, j^*+k]$ in the finetuning process. The perturbation is achieved by forwarding the sliding window toward the right first.  When the MMD~\cite{mmd} decreases, the perturbation direction will be kept in the interval, and be reversed when the MMD~\cite{mmd} increases. In this way, the model is able to converge in the optimal frequency band and achieves the best transferability. The whole algorithm of our proposed frequency movement-based alignment is shown in Alg.~\ref{alg::frequency movement}. $k$ is set as 3 in our paper. 
 
 \begin{algorithm}[htp]
   \caption{Frequency Movement-based Alignment}
   \label{alg::frequency movement}
   \begin{algorithmic}[1]
    \State \textbf{Input:}
    $\widetilde{\mathcal{F}}_{s}^{j}$, $\widetilde{\mathcal{F}}_{t}^{j}$:
     the $j^{th}$ frequency bands of the source and target domains; $m$: the number of frequencies in one window;  $j$: current move step; $T, T_{m}, T_{w}, t$: iterations of the $j^{th}$ movement interval, frequency movement, warm-up, and current iterations. 
     \State \textbf{Init:} Parameters $\phi=\{\theta_G, \theta_D, \theta_R\}$ of generator, discriminator and regression head.   
     \While{$t < T_{w}$}  {\textcolor{blue}{\Comment{Warm up with pre-training}}}  
       \State  pre-training the $\phi$ by aligning the $\widetilde{\mathcal{F}}_{s}^{0}$, and $\widetilde{\mathcal{F}}_{t}^{0}$.
     \EndWhile
     \While{$T_{w}<=t<T_{m}$}  {\textcolor{blue}{\Comment{Frequency movement-based selection}}}  
       \State Move sliding window forward along the pre-defined trajectory: $j\xleftarrow{}j+1$
       \State Align the $\widetilde{\mathcal{F}}_{s}^{j}$ and $\widetilde{\mathcal{F}}_{t}^{j}$, and obtain the transferability $\overline{\epsilon}_j$ for $T$ iterations, $t\xleftarrow{}t+T$.
     \EndWhile 
     \State Obtain initial optimal frequency band
     \State \quad \quad $j^*{^{th}}=argmax[\overline{\epsilon}_0, \overline{\epsilon}_1, ..., \overline{\epsilon}_T]$ 
     \State \textbf{Init: $d=+1$, $j=j^*$}
     \While{$T_{m}<=t<T_{a}$}  {\textcolor{blue}{\Comment{Perturbation-based finetuning}}}  
         \State Perturbed the frequency band with sliding window by $j\xleftarrow{}j+d$. 
         \State Finetuning the $\phi$ by aligning $\widetilde{\mathcal{F}}_{s}^{j}$ and $\widetilde{\mathcal{F}}_{t}^{j}$, and obtain the transferability value $\overline{\epsilon}_{j}$.
         \If{$\overline{\epsilon}_{j}>\overline{\epsilon}_{j-d}$ or $j>=j^{*}+k$ or $j<=j^{*}-k$}
         \State $d\xleftarrow{}(-d)$  {\textcolor{blue}{\Comment{Reverse direction}}}   
         \EndIf
     \EndWhile
\end{algorithmic}
 \end{algorithm}

\subsubsection{Frequency Alignment}
\tcr{In the process of frequency movement, we exploit the frequency alignment to enable the BIQA model to achieve powerful perception transferability. Specifically, one discriminator $D$ is employed to distinguish the source and target domains for selected frequencies. Here, we denote the selected frequencies as $\Tilde{\mathcal{F}}_{s}^*$ and $\Tilde{\mathcal{F}}_{t}^*$. The adversarial-based feature alignment~\cite{DANN1} is described as: 
\begin{equation}
    \mathcal{L}_{adv} = - \mathbb{E}[log(1- D(\Tilde{\mathcal{F}}_{s}^*))-\mathbb{E}[log(D(\Tilde{\mathcal{F}}_{t}^*))]
   \label{equ:adv_loss}
\end{equation}
 With the help of  the gradient reversal layer~\cite{UDA_GRL}, we can achieve end-to-end adversarial-based feature alignment by multiplying gradients with a negative constant during back-propagation to ensure that the frequency space on the two domains is indistinguishable.}

\subsection{Overall Optimization Losses}
\label{D}
To boost the perception transferability of the BIQA model, we introduce three losses in our method, \ieno, adversarial-based feature alignment loss $\mathcal{L}_{adv}$, regression loss $\mathcal{L}_S$ and self-supervised loss $\mathcal{L}_T$.
\begin{table*}[]
\centering

\caption{The details of the commonly-used datasets in BIQA, where ``$\uparrow$” means higher score for higher image quality, $\downarrow$ is the opposite.}
\label{tab:data}
\setlength{\tabcolsep}{3mm}{
\begin{tabular}{c|lcclcc}
\hline
Scenario                   & Database       & \multicolumn{1}{l}{Dist. images} & \multicolumn{1}{l}{Dist. Types} & Annotation     & \multicolumn{1}{l}{Score Range} & Score Meaning \\ \hline
\multirow{2}{*}{Synthetic} & KADID10k~\cite{Kadid10k}       & 10125                            & 25                              & MOS, variance  & {[}1,5{]}                       & ~$\uparrow$    \\ \cline{2-7} 
                           & CSIQ~\cite{CSIQ}           & 866                              & 6                               & DMOS, variance & {[}0,1{]}                       & ~$\downarrow$   \\ \hline
\multirow{2}{*}{Authentic} & KonIQ-10k~\cite{KonIQ}      & 10073                            & -                               & MOS, variance  & {[}1,5{]}                       & ~$\uparrow$    \\ \cline{2-7} 
                           & CLIVE~\cite{LIVEC}          & 1162                             & -                               & MOS, variance  & {[}0,100{]}                     & ~$\uparrow$    \\ \hline
Realistic Blur             & BID~\cite{BID} & 586                              & -                               & MOS, variance  & {[}0,5{]}                       & ~$\uparrow$     \\ \hline
\end{tabular}}
\end{table*}
\begin{table*}[]
\centering
\caption{\tct{The performance comparison with the state-of-the-art BIQA methods on the synthetic to authentic domains. The best performance is bold in this table.}}
\label{tab:compare1}
\setlength{\tabcolsep}{0.7mm}{
\begin{tabular}{c|cccccc|c}
\hline \hline
              & \begin{tabular}[c]{@{}c@{}}KADID10k$\rightarrow$KonIQ-10k\\ SROCC/PLCC\end{tabular} & \begin{tabular}[c]{@{}c@{}}KADID10k$\rightarrow$CLIVE\\   SROCC/PLCC\end{tabular} & \begin{tabular}[c]{@{}c@{}}CSIQ$\rightarrow$KonIQ-10k\\ SROCC/PLCC\end{tabular} & \begin{tabular}[c]{@{}c@{}}CSIQ$\rightarrow$CLIVE\\  SROCC/PLCC\end{tabular} &
              \begin{tabular}[c]{@{}c@{}}KADID10k$\rightarrow$BID\\  SROCC/PLCC\end{tabular} &
              \begin{tabular}[c]{@{}c@{}}CSIQ$\rightarrow$BID\\  SROCC/PLCC\end{tabular} &
              \begin{tabular}[c]{@{}c@{}}Average\\  SROCC/PLCC\end{tabular}\\ \hline 
NIQE~\cite{NIQE}          & 0.4469/0.4600                                                             &   0.3044/0.3619                                                           & 0.4469/0.4600  
                          & 0.3044/0.3619   
                          &0.3553/0.3812 
                          &0.3553/0.3812
                          & 0.3688/0.4010
                          \\ 
PIQE~\cite{PIQE}          & 0.0843/0.1995                                                             &0.2622/0.3617 
                           & 0.0843/0.1995                                         
                           & 0.2622/0.3617 
                           &0.2693/0.3506 
                           &0.2693/0.3506
                           & 0.2052/0.3039\\ 
BRISQUE~\cite{BRISQUE}       & 0.1077/0.0991                                                             &  0.2433/0.2512                                                            & 0.2336/0.2833  
                             &0.1736/0.2195
                             &0.1745/0.1750 
                             &0.2216/0.2810
                             &0.1923/0.2181 \\ \hline \hline
DBCNN~\cite{DBCNN}         & 0.4126/0.4209                                                             & 0.2663/0.2897  
                           & 0.3905/0.3186
                           & 0.4088/0.3619   
                           &0.3179/0.2115 
                           &0.4854/0.4734
                           &0.3802/0.3459  \\ 
HyperIQA~\cite{HyperIQA}      & 0.5447/0.5562                                                             & 0.4903/0.4872                                                             & 0.5501/0.5402                                                             & 0.3687/0.3676
                              &0.3794/0.2820 
                              &0.5611/0.5298
                              &0.4823/0.4604 \\ 
RankIQA~\cite{Rankiqa}       & 0.6030/0.5511                                                             & 0.4906/0.4950                                                             & 0.5665/0.5424  
                             & 0.4705/0.4546
                             &0.5101/0.3671 
                             &0.5961/0.5613
                             &0.5394/0.4952\\
MUSIQ~\cite{MUSIQ}          & 0.5541/0.5732                                                             &0.5171/0.5243
                           & 0.5748/0.5996                                         
                           & 0.4309/0.4770
                           &0.5070/0.5312 
                           &0.6059/0.5448
                           & 0.5316/0.5416\\ 
GraphIQA~\cite{sun2022graphiqa}       & 0.4268/0.4297                                                            & 0.3884/0.4070                                                            & 0.5250/0.5325 
                             &0.4710/0.4799
                             &0.3868/0.1897
                             &0.5598/0.5363
                             & 0.4596/0.4291\\ 
VCRNet~\cite{pan2022vcrnet}         & 0.5661/0.5848                                                            & 0.5197/0.5302 
                           & 0.5443/0.5454
                           & 0.4645/0.4824   
                           &0.4719/0.3650
                           &0.5420/0.5174
                           & 0.5180/0.5042\\

  \hline \hline 

DANN~\cite{DANN1}          & 0.6382/0.6360                                                             & 0.4990/0.4835                                                             & 0.6192/0.6186
                           & 0.5265/0.5174
                           &0.5861/0.5102 
                           &0.5722/0.5107
                           & 0.5735/0.5460 \\ 
UCDA~\cite{UCDA}          & 0.4958/0.5010                                                             & 0.5830/0.6192                                                             &0.5613/0.5947                 
                         &0.5041/0.5476
                         &0.3480/0.3907 
                         &0.4898 /0.4993
                         & 0.4969/0.5254\\ 
RankDA~\cite{RankDA}     & 0.6383/0.6227                                                             & 0.4512/0.4548                                                             & 0.5831/0.5749                                                             & 0.2789/0.2323
                         &0.5350/0.5820 
                         &0.3683/0.3177
                         &0.4758/0.4640\\ 

StyleAM~\cite{lu2022styleam} & 0.7002/0.6733                                                             & 0.5844/0.5606                                                             & 0.6483/0.6070                                                             & 0.5061/0.5416
                             &0.6365/0.5669 
                             &0.6615/0.6070
                             &0.6228/0.5927\\ 
SFUDA~\cite{SUFDA}   & 0.7221/0.7124                                                             & 0.5562/0.5673
                     & \textbf{0.6669}/0.6744
                     & 0.4477/0.5060
                     &0.5946/0.5982 
                     &0.5935/0.5996
                     &0.5968/0.6096\\ \hline \hline

Baseline     & 0.6346/0.5946    
              & 0.4959/0.5020                                               & 0.4662/0.4941                                              & 0.4550/0.4649
              &0.5600/0.5200 
              &0.5518/0.5018
              &0.5272/0.5129\\ 
\rowcolor[gray]{0.9} FreqAlign(ours)   & \textbf{0.7478/0.7213}
                  & \textbf{0.6175/0.5884}
                  & \textbf{0.6743/0.7082}
                  & \textbf{0.5540/0.5852}
                   &\textbf{0.6329/0.6249} 
                  &\textbf{0.6896/0.7077}
                  & \textbf{0.6526/0.6559} \\ \hline

\end{tabular}}
\end{table*}
\tcr{Here, the self-supervised loss $\mathcal{L}_{T}$ on target domain is proposed by SFUDA~\cite{SUFDA}, which has revealed that their proposed self-supervised loss is able to improve the  source-free UDA of BIQA in conjunction with the characteristic of BIQA. Inspired by this, we further employ the self-supervised loss function to constrain the distribution of target data and learn the discriminative information of the target domain.}

\section{Experiments}
\subsection{Datasets}
\label{sec:dataset}
\tcr{We evaluate our proposed scheme on tree typical cross-domain scenarios, including synthetic dataset to authentic dataset, authentic dataset to synthetic dataset, and cross-distortion scenarios in the UDA of BIQA. In the above three settings, two commonly-used synthetic datasets, \ieno, KADID10k~\cite{Kadid10k}, 
CSIQ~\cite{CSIQ} and three authentic datasets are adopted, including KonIQ-10k~\cite{KonIQ}, CLIVE~\cite{LIVEC}, and one realistic blur dataset BID~\cite{BID}. As described in Table~\ref{tab:data}, Kadid10k contains 25 synthetic distortions, which can be divided into 7 categories: blur, color distortion, compression, noise, brightness change, spatial distortion, sharpness, and contrast. CSIQ collects 6 distortion types, including JPEG compression, JPEG-2000 compression, global contrast decrements, additive pink Gaussian noise, and Gaussian blurring. KonIQ-10k and CLIVE are composed of 10073 and 586 images with complicated realistic distortions. Following~\cite{lu2022styleam}, each distortion category in the KADID10k~\cite{Kadid10k} as each domain for the cross-distortion scenarios. And we exploit the commonly-used leave-one-domain protocol for the cross-distortion scenario.}

\tcr{To evaluate the performance of our method in improving the perception transferability of BIQA, we utilize two synthetic datasets and three authentic datasets to construct $2 \times 3$ synthetic domain to authentic domain settings and $3 \times 2$ authentic domain to synthetic domain settings. Notably,  we also investigate the effects of our method on pure cross-distortion settings. Concretely, seven distortion types in KADID10k~\cite{Kadid10k} construct $7\times 1$ cross-distortion settings, which adopt one distortion type as the target domain and others as source domains. To unify the quality range of different datasets, the annotated quality labels are rescaled linearly to the range of [1, 5] following~\cite{lu2022styleam}. In the training process, each image is randomly cropped into $384\times 384$. We utilize random horizontal flipping as our data augmentation strategy. In the inference stage, all images are cropped into $384\times384$ at the center. 
}

We employ the ResNet18 pretrained on ImageNet~\cite{imagenet} together with two fully connected layers (FC) and ReLU layers (\ieno, FC-ReLU-FC-ReLU)  as our backbone. For the domain discriminator, we exploit the architecture of two linear layers and one sigmoid function (\ieno, FC-ReLU-FC-ReLU-Sigmoid). The regression head is composed of one FC layer and one Softmax layer. All experiments are implemented with  Pytorch~\cite{pytorch} and are trained with one NVIDIA GeForce 1080Ti GPU. The model is optimized with a batch size of 40 for 300 epochs.  In this process, we utilize Adam optimizer~\cite{kingma2014adam} with a learning rate of $1e^{-4}$ and a weight decay of $5\times e^{-4}$ for optimization.

\begin{table*}[]
\centering
\caption{\tct{The performance comparison with the state-of-the-art BIQA methods on  authentic to synthetic domains. The best performance is bold.}}
\label{tab:compare2}
\setlength{\tabcolsep}{0.5mm}{
\begin{tabular}{c|cccccc|c}
\hline\hline
              & \begin{tabular}[c]{@{}c@{}}KonIQ-10k$\rightarrow$KADID10k\\ SROCC/PLCC\end{tabular} & \begin{tabular}[c]{@{}c@{}}CLIVE$\rightarrow$KADID10k\\   SROCC/PLCC\end{tabular} & \begin{tabular}[c]{@{}c@{}}KonIQ-10k$\rightarrow$CSIQ\\ SROCC/PLCC\end{tabular} & \begin{tabular}[c]{@{}c@{}}CLIVE$\rightarrow$CSIQ\\  SROCC/PLCC\end{tabular} &
              \begin{tabular}[c]{@{}c@{}}BID$\rightarrow$KADID10k\\  SROCC/PLCC\end{tabular} &
              \begin{tabular}[c]{@{}c@{}}BID$\rightarrow$CSIQ\\  SROCC/PLCC\end{tabular} &
              \begin{tabular}[c]{@{}c@{}}Average\\  SROCC/PLCC\end{tabular}\\ \hline
NIQE~\cite{NIQE}          & 0.1010/0.1541                                                             & 0.1010/0.1541                                                             & 0.2996/0.4059
                          &0.2996/0.4059 
                          &0.1010/0.1541 
                          &0.2996/0.4059
                          &0.2003/0.2799   \\  
PIQE~\cite{PIQE}          &0.2846/0.3607                                                             & 0.2846/0.3607
                           &0.6239/0.7165                                &0.6239/0.7165 
                           &0.2846/0.3607   
                           &0.6239/0.7165
                           & 0.4542/0.5386 \\ 
BRISQUE~\cite{BRISQUE}       &0.0881/0.2006                                                           & 0.1510/0.1887                                                                &  0.2430/0.29205 
                             &0.0955/0.1648
                             &0.0881/0.2006 
                             &0.2430/0.2920
                             &0.1514/0.2231\\ \hline \hline
DBCNN~\cite{DBCNN}         & 0.4549/0.4635                                                              &0.4194/0.4186
                           & 0.6447/0.5506
                           & 0.6092/0.6123
                           &0.4579/0.4367 
                           &0.6384/0.5984
                           & 0.4947/0.4906 \\ 
HyperIQA~\cite{HyperIQA}      & 0.5114/0.5163                                                             & 0.3797/0.4391                                                         & 0.5975/0.5675                                                             & 0.5592/0.5616
                          &0.4131/0.4282 
                          &0.5078/0.4315
                              &0.4941/0.4676\\ 
RankIQA~\cite{Rankiqa}       &  0.4873/0.4255                                                            & 0.4158/0.3904                                                              & 0.5791/0.5392
                             &0.5114/0.5062
                             &0.4137/0.4404 
                             &0.5573/0.5043
                             &0.5162/0.5270\\
MUSIQ~\cite{MUSIQ}          &0.4861/0.5042                                                             & \textbf{0.4969/0.5238}
                           &0.5294,/0.5578                               &\textbf{0.7285/0.7554} 
                           & 0.3489/0.4022  
                           &0.6511/0.6992
                           & 0.5401/0.5737\\ 
GraphIQA~\cite{sun2022graphiqa}       &0.5215/0.5360                                                           & 0.3647/0.3925                                                                &  0.6501/0.6679
                             &0.5234/0.5471
                             &0.3918/0.4004 
                             &0.5519/0.4953
                             &0.5005/0.5065 \\
VCRNet~\cite{pan2022vcrnet}         & 0.5038/0.5165                                                              &0.3404/0.4160
                           & 0.5969/0.5737
                           & 0.6325/0.6215
                           &0.2379/0.1685 
                           &0.3293/0.3479
                           & 0.4401/0.4406 \\
                             \hline \hline

DANN~\cite{DANN1}          &0.5435/0.5523
                           &0.3829/0.4321     
                           & 0.6192/0.6186
                           &0.7219/0.7293 
                           &0.4385/ 0.4378 
                           &0.5258/0.4470
                           &0.5386/0.5361 \\ 
UCDA~\cite{UCDA}           & 0.4297/0.4372                                                              &0.4198/0.4850                                                            & 0.6214/0.6279                                                              & 0.6098/0.6491
                           &0.2656/0.3261 
                           &0.5088/0.5248
                           &0.4758/0.5083\\ 
RankDA~\cite{RankDA}     &  0.4031/0.4130                                                             & 0.3773/0.4433                                                              & 0.5837/0.5731                                                               & 0.2750/0.3040
                           &0.3227/0.1894 
                           &0.5093/0.4822
                         &0.4118/0.4008\\ 

StyleAM~\cite{lu2022styleam} &0.5701/0.6008                                                              & 0.4080/0.4705                                                            & 0.7157/0.6818                                                              & 0.7012/0.7178
                             &0.4521/0.4972 
                             &\textbf{0.6883/0.7081}
                             & 0.5892/0.6126\\ 
SFUDA~\cite{SUFDA}   & 0.5829/0.6024                                                             & 0.2532/0.4127
                     & 0.7699/0.8121 
                     & 0.3332/0.3844
                     &0.3960/0.4606 
                     &0.4743/0.4432
                     & 0.4682/0.5192 \\ \hline \hline
Baseline      &  0.5526/0.5402   
              &  0.3757/0.4490                                       
              &  0.6218/0.5719                                           
              &  0.6382/0.6215
              &0.3757/0.4491 
              &0.5337/0.5305
              &0.5162/0.5270\\
\rowcolor[gray]{0.9} FreqAlign (ours)   & \textbf{0.5968/0.5798}
                  & 0.4313/0.4983
                  & \textbf{0.8044/0.7936}
                  &0.6897/0.6918
                  &\textbf{0.4799/0.5103} 
                  &0.6224/0.6102
                  & \textbf{0.6040/0.6140} \\ \hline 
\end{tabular}}
\end{table*}

\begin{table*}[]
\centering
\caption{\tct{The performance comparison with the state-of-the-art BIQA methods on cross distortion types in KADID10k, where ``X/X” means ``SROCC/PLCC” and types 1-7 represents blur, color distortion, compression, noise, brightness change, spatial distortion, sharpness, and contrast, respectively. The best performance is bold.}}
\label{tab:cross_dis}
\setlength{\tabcolsep}{1mm}{
\begin{tabular}{c|ccccccc|c}
\hline 
        Methods               & Others$\rightarrow$Type1              & Others$\rightarrow$Type2   & 
                       Others$\rightarrow$Type3            & Others$\rightarrow$Type4      & Others$\rightarrow$Type5 & 
                       Others$\rightarrow$Type6 & 
                       Others$\rightarrow$Type7 &
                       Average \\ \hline
NIQE~\cite{NIQE}                   & 0.4263/0.5597  & 0.1080/0.1642  & 0.2460/0.2654   & 0.3027/0.3190   & 0.3187/0.5605    & 0.1424/0.1586    & 0.3102/0.3470  &0.2649/0.3392\\ 
PIQE~\cite{PIQE}                   & 0.6785/0.6909  & 0.0987/0.1892 & 0.7113/0.7808   & 0.1836/0.2764    & 0.2945/0.4304 &  0.0248/0.0354 &  0.3695/0.3181 & 0.3372/0.3890\\ 
BRISQUE~\cite{BRISQUE}                   &0.0037/0.2171   & 0.2743/0.3488 &  0.1080/0.1313  & 0.0039/0.0663   & 0.2041/0.3577 & 0.0027/0.1121 & 0.1683/0.1318 & 0.1092/0.1950\\ \hline \hline
DBCNN~\cite{DBCNN}                  &0.8218/0.7503   &0.2828/0.2221  & 0.8448/0.8938   & 0.8094/0.7866   &\textbf{0.5900/0.6869}  & 0.4415/0.4420 &0.6924/0.7344  &0.6403/0.6451\\ 
HyperIQA~\cite{HyperIQA}               &0.5296/0.4981   &0.3016/0.2536  & 0.8667/0.9038   & 0.8247/\textbf{0.8202}   &0.4852/0.6638  & 0.3533/0.4008  & 0.7321/0.7839 &0.5847/0.6177 \\ 
RankIQA~\cite{Rankiqa}                  & 0.7352/0.7059  &0.3516/0.3642  &0.8168/0.8079    & 0.7849/0.7735   &0.3465/0.5528  & 0.3348/0.4118 & 0.5917/0.6408 & 0.5685/0.6081\\
MUSIQ~\cite{MUSIQ}                   &0.7865/0.7122   &0.3543/0.3760   &0.7953/0.8651    &0.7436/0.7433    &0.5029/0.5337     &\textbf{0.5297/0.5534}    & 0.4626/0.5048  &0.5964/0.6126\\ 
GraphIQA~\cite{sun2022graphiqa}                   &0.7991/0.7454   &0.4417/0.4469  & 0.8072/0.8189   &0.6201/0.6104   &0.3774/0.5518  &0.3610/0.3926  &0.5159/0.5933   & 0.5603/0.5941\\ 
VCRNet~\cite{pan2022vcrnet}                   &0.5760/0.5609  &0.3866/0.3996  &0.8525/0.8774    & 0.7371/0.7248   &0.4292/0.5684 &0.3630/0.4173  &0.7003/0.7501  &0.5778/0.6140 \\\hline \hline

DANN~\cite{DANN1}                        & 0.6222/0.5835  &0.5273/0.4930  &  0.7980/0.82102   & 0.7389/0.7190   &  0.4734/0.5449  &0.3675/0.4004 & 0.4830/0.5009 & 0.5729/0.5802\\ 
UCDA~\cite{UCDA}                      & 0.6043/0.5827  & 0.2885/0.3424 & 0.8430/\textbf{0.9132}  &0.6199/0.6365    & 0.2364/0.3935 & 0.3808/0.4404 &  0.4534/0.5248& 0.4894/0.5480\\ 
RankDA~\cite{RankDA}                       & 0.6253/0.5739  &0.5690/0.5557  &0.4106/0.2734    & 0.4695/0.4030   & 0.3158/0.4073 & 0.1958/0.1851 & 0.6765/0.6772 &0.4712/0.4393\\ 
      

StyleAM~\cite{lu2022styleam}                & 0.8277/0.8291                        & 0.4752/0.4719                           & 0.8763/0.9097                           & \textbf{0.8340}/0.8199       & 0.5120/0.6482                      & 0.4731/0.3628                         & 0.7472/0.7023                       &  0.6765/0.6777   \\ 
SFUDA~\cite{SUFDA}                & \textbf{0.9498/0.9591}                         
                                  & 0.2929/0.3577                         
                                  & 0.8072/0.8669                           
                                  & 0.8155/0.8166       
                                  &0.4002/0.5875                    
                                  &0.3888/0.4603                        
                                  &0.4927/0.5461                        
                                  &0.5924/0.6563 \\ \hline \hline

Baseline               & 0.5377/0.5171                          & 0.4440/0.4412                           & 0.8207/0.8567                               & 0.8150/0.8088       & 0.4841/0.5795               & 0.4121/0.4087                 & 0.5274/0.5418       &  0.5773/0.5934             \\ 
\rowcolor[gray]{0.9} FreqAlign    & 0.8905/0.8730 
                                  & \textbf{0.6005/0.5958}
                                  & \textbf{0.8957}/0.8648                           
                                  & 0.7872/0.7615    
                                  & 0.4597/0.5843
                                  & 0.5094/0.3490                       
                                  & \textbf{0.7624/0.7850}   
                                  & \textbf{0.7007/0.6876}  \\ \hline

\end{tabular}}
\end{table*}

\tcr{We adopt two standard criteria to measure the performance of BIQA, \ieno, Spearman
rank-order correlation coefficient (SROCC) and Pearson linear correlation coefficient (PLCC), which are good at measuring the monotonicity
and precision of predicted quality scores, respectively. With the increase of SROCC/PLCC, the BIQA model is more effective for BIQA. For PLCC, the predicted quality scores require to be fitted through a non-linear logistic mapping function:
\begin{equation}
    \hat{y}={{\beta }_{1}}(\frac{1}{2}-\frac{1}{\exp ({{\beta }_{2}}(\hat{y}-{{\beta }_{3}}))})+{{\beta }_{4}}\hat{y}+{{\beta }_{5}}
\end{equation}
}

\subsection{Performance Evaluation}

\tcr{We validate the effectiveness of our FreqAlign on three typical cross-domain scenarios, which includes 3$\times$2 ``synthetic $\to$ authentic” cross-domain settings, 2$\times$3 ``authentic $\to$ synthetic” cross-domain settings and $7\times 1$ cross-distortion settings.  The baseline is set as the pretrained ResNet-18~\cite{ResNet}, which is trained only with the source domain data and directly tested on the target domain. We select three categories of BIQA methods for comparison: 1) three state-of-the-art (SOTA) traditional BIQA methods, including NIQE~\cite{NIQE}, BRISQUE~\cite{BRISQUE} and PIQE~\cite{PIQE}, 2) six SOTA learning-based works, \ieno, RankIQA~\cite{Rankiqa}, DBCNN ~\cite{DBCNN}, HyperIQA~\cite{HyperIQA}, \tct{MUSIQ~\cite{MUSIQ}, GraphIQA~\cite{sun2022graphiqa} and VCRNet~\cite{pan2022vcrnet}}, 3) four SOTA works for the UDA of BIQA, including UCDA~\cite{UCDA}, RankDA~\cite{RankDA}, StyleAM~\cite{lu2022styleam} and SFUDA~\cite{SUFDA}. Particularly,
\textbf{UCDA}~\cite{UCDA} employs the feature of the last layer in ResNet18 to achieve domain alignment. \textbf{RankDA}~\cite{RankDA} adopt the rank features of distorted image pair to transfer perceptional rank information from the source domain to the target domain. In \textbf{SFUDA}~\cite{SUFDA}, three self-supervised losses are introduced based on the quality distribution of human annotation, which behaviors excellent performances on the source-free domain adaptation of BIQA.  \textbf{StyleAM} aims to find the perception space existing in the features of BIQA, and introduce perception-oriented feature alignment for the UDA of BIQA.
All the UDA works for BIQA share the same backbone ResNet18~\cite{ResNet} as our FreqAlign.}

\subsubsection{Synthetic $\rightarrow$ Authentic Domains}
The experimentally quantitative comparisons are shown in Table~\ref{tab:compare1}. We systematically investigate our FreqAlign on six cross-domain settings, including Kadid10k $\rightarrow$ KonIQ-10k, Kadid10k $\rightarrow$ CLIVE, Kadid10K$\rightarrow$ BID, CSIQ $\rightarrow$ KonIQ-10k, CSIQ $\rightarrow$ CLIVE, and CSIQ$\rightarrow$ BID. From the table, we can have the following conclusions: 1) Existing traditional and learning-based methods for BIQA perform poorly on unseen domains. Despite the DBCNN~\cite{DBCNN} and HyperIQA~\cite{HyperIQA} having been designed for authentic distortion, 
they still fail when meeting the unseen authentic dataset. Existing traditional and learning-based methods for BIQA perform poorly on unseen domains. Despite the DBCNN~\cite{DBCNN} and HyperIQA~\cite{HyperIQA} having been designed for authentic distortion, 
they still fail when meeting the unseen authentic dataset. \tct{MUSIQ~\cite{MUSIQ}, GraphIQA~\cite{sun2022graphiqa}, and VCRNet~\cite{pan2022vcrnet} also perform poor generalization capability on our cross-domain settings.} 2) Our proposed FreqAlign achieves the optimal performance on PLCC for all six cross-domain settings. Compared with DANN~\cite{DANN1}, which aligns the source and target domain at the low-frequency component, ours improves it by a large margin of 0.0791/0.1099 on average SROCC/PLCC. This reveals the importance and effectiveness of our proposed FreqAlign. 3) It is noteworthy that our FreqAlign is clearly superior to recent UDA-based BIQA works without any complicated operations. Concretely, on the challenging cross-domain setting CSIQ $\rightarrow$ CLIVE, ours outperform the second optimal UDA method StyleAM~\cite{lu2022styleam} by \tcb{0.0479/0.0436} on SROCC/PLCC.

\begin{table*}[]
\caption{The ablation studies for the frequency movement-based alignment.}
\label{tab:all_ablation}
\begin{tabular}{cc|c|c|c|c|c|c}
\hline
\multicolumn{2}{c|}{Frequency Movement}                    & \multirow{2}{*}{$\mathcal{L}_T$} & \multicolumn{1}{c|}{KADID10k$\rightarrow$KonIQ-10k} & KADID10k$\rightarrow$CLIVE & CSIQ$\rightarrow$ KonIQ-10K & CSIQ$\rightarrow$CLIVE & \multirow{2}{*}{Average} \\ \cline{1-2}
\multicolumn{1}{l|}{Warm up} & Perturbation &                       & \multicolumn{1}{c|}{SROCC/PLCC}                       & \multicolumn{1}{c|}{SROCC/PLCC}  & \multicolumn{1}{c|}{SROCC/PLCC}  & \multicolumn{1}{c|}{SROCC/PLCC}  &                          \\ \hline
\multicolumn{1}{c|}{\XSolidBrush}        &    \XSolidBrush       & \multirow{4}{*}{\CheckmarkBold}     & 0.6995/0.6797                                                       & \tcb{0.5653/0.5732}                                 & 0.6573/0.6822                                &    0.4821/0.5378                              &0.6010/0.6182                          \\ \cline{1-2} \cline{4-8} 
\multicolumn{1}{c|}{\CheckmarkBold}        &     \XSolidBrush        &                       &  0.7235/0.7047                                                     &    \tcb{0.5790/0.5806}                              &   0.6708/0.6928                               &  0.4952/0.5340                                & 0.6171/0.6280                 \\ \cline{1-2} \cline{4-8} 
\multicolumn{1}{c|}{\CheckmarkBold}        &    \CheckmarkBold         &                       &    0.7478/0.7213                                                   &    0.6175/0.5884                              &       0.6743/0.7082                          &  0.5540/0.5852                                & 0.6484/0.6507                     \\ \hline
\multicolumn{1}{c|}{\CheckmarkBold}        &   \CheckmarkBold          &     \XSolidBrush                  &       0.6788/0.6777                                                &  0.5354/0.5530                                &   0.6384/0.6580                               &    0.5205/0.5558                             &   0.5933/0.6111                       \\ \hline
\multicolumn{2}{c|}{Gumbel-based}          &   \CheckmarkBold                    &     0.6947/0.6824                                                  &     0.5682/0.5926                              &    0.6620/0.6777                              &      0.5223/0.5678                             &  0.6118/0.6301                        \\ \hline
\end{tabular}
\end{table*}

\begin{table}[]
\centering
\caption{A comparison between three different trajectories for frequency movement, including row, column, and zigzag.}
\label{tab:trajectory}
\begin{tabular}{c|cc}
\hline
  \begin{tabular}[c]{@{}c@{}}Trajectory\end{tabular} 

  & \begin{tabular}[c]{@{}c@{}}KonIQ-10k$\rightarrow$KADID10k\\   SROCC/PLCC\end{tabular} 

  & \begin{tabular}[c]{@{}c@{}}Kadid10k$\rightarrow$KonIQ-10k\\   SROCC/PLCC\end{tabular} 
 \\ \hline

Left-to-right   
& 0.5968/0.5798                                                                       
& 0.7478/0.7213                                                                                                                           
\\ \hline
Up-to-down                                                                      
& 0.5774/0.5928                                                                
& 0.6933/0.6924                                                            
\\ \hline

Zig-zag Scanning                                                                          
& 0.6008/0.6064                                                                  
&  0.7212/0.7220                                                        
\\ \hline
\end{tabular}
\end{table}

\subsubsection{Authentic $\rightarrow$ Synthetic Domains}

Depart that our FreqAlign has achieved great performance on synthetic to authentic domain settings, we also explore the effectiveness of our FreqAlign on the synthetic to  authentic domain  settings. Concretely, we adopt three authentic datasets (\ieno, KonIQ-10k, CLIVE, and BID) as the source domains and two synthetic datasets (\ieno, KADID10k, and CSIQ) as target domains. The experimental results are described in Table~\ref{tab:compare2}. From the table, we can observe that our FreqAlign still obtains excellent perception transferability on these authentic to synthetic settings. Moreover, ours outperforms the baseline by an average of \tcb{0.0878/0.0870} on SROCC/PLCC, and surpasses the second optimal method StyleAM~\cite{lu2022styleam} by \tcb{0.0148} on SROCC.

\begin{figure}
    \centering
\includegraphics[width=1\linewidth]{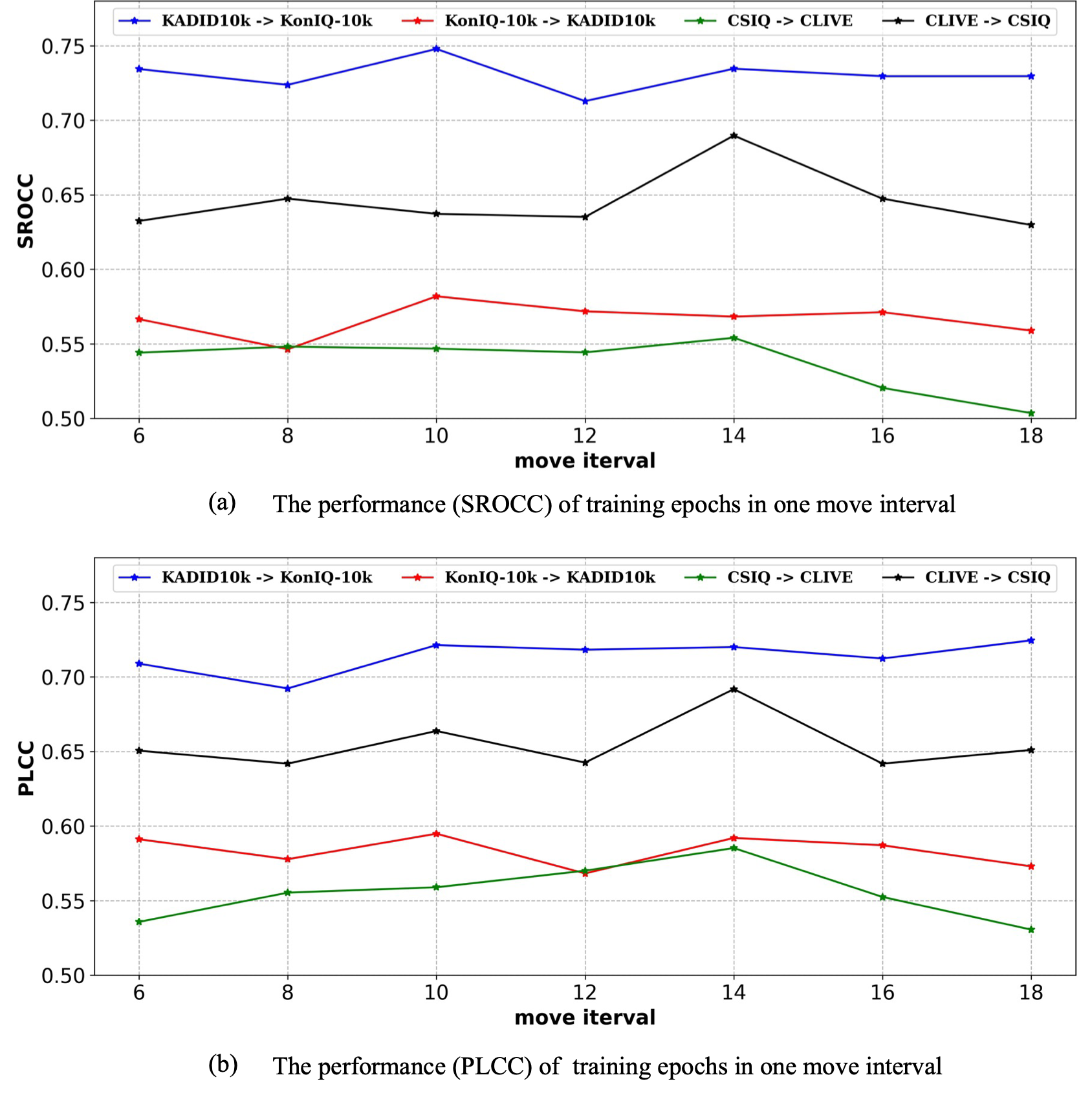}
    \caption{\tct{The performance curve of different move intervals.}}
    \label{fig:ablation_for_interval_group}
\end{figure}
\subsubsection{Cross Distortion Types}
\tcr{We notice that the distribution shifts of the above cross-domains settings are mostly caused by the differences in contents and distortions. To verify the performance of our algorithm on the pure cross-distortion settings, we select the KADID10k dataset that contains seven distortion types for experiments. In particular, we follow the multiple-source domain adaptation~\cite{peng2019moment} and utilize the leave-one-domain protocol. We report the performances in Table~\ref{tab:cross_dis}. From the Table, we can find most UDA works of BIQA perform poorly in some specific domains. For instance, UCDA~\cite{UCDA} and SFUDA~\cite{SUFDA}  suffer from poor performance with the SROCC of 0.2885 and 0.2929 on color distortions. RankDA~\cite{RankDA} obtains limited performance on the sharpness distortion with SROCC of 0.1958. In contrast, our FreqAlign shows stable and remarkable perception transferability on all seven distortion types. Furthermore, our FreqAlign outperforms the second optimal UDA method StyleAM by a large margin of \tcb{0.0242} in terms of SROCC.}

\subsection{Ablation Study}
\begin{table*}[]
\centering
\caption{A comparison between three different metrics for the guidance of frequency movement, including MMD~\cite{mmd}, CORAL~\cite{CORAL}, and adversarial loss~\cite{DANN1} (\ieno, Adv. Loss in this table).}
\label{tab:ab_metric}
\begin{tabular}{c|cccc|c}
\hline
  \begin{tabular}[c]{@{}c@{}}Transferability\\   Metric\end{tabular} 
  & \begin{tabular}[c]{@{}c@{}}CSIQ$\rightarrow$KonIQ-10k\\  SROCC/PLCC\end{tabular} 
  & \begin{tabular}[c]{@{}c@{}}KonIQ-10k$\rightarrow$KADID10k\\   SROCC/PLCC\end{tabular} 
  & \begin{tabular}[c]{@{}c@{}}CSIQ$\rightarrow$CLIVE\\  SROCC/PLCC\end{tabular} 
  & \begin{tabular}[c]{@{}c@{}}Kadid10k$\rightarrow$KonIQ-10k\\   SROCC/PLCC\end{tabular} 
  &\begin{tabular}[c]{@{}c@{}}average\\  SROCC/PLCC\end{tabular} \\ \hline

MMD~\cite{mmd}      
& 0.6743/0.7082                                                                       
& 0.5968/0.5798                                                                    
&0.5540/0.5852                                                                   
& 0.7478/0.7213
                                                                   
&0.6432/0.6486\\ \hline
CORAL ~\cite{CORAL}                                                                        
&0.6835/0.7099                                                                        
&0.5640/0.5782                                                                    
&0.5250/0.5795                                                                       
& 0.7286/0.7123
                                                                  
&0.6252/0.6449 \\ \hline

Adv. Loss ~\cite{DANN1}                                                                           
&0.6661/0.6952                                                                     
&0.5696/0.5817                                                                    
& 0.5340/0.5749                                                                       
& 0.7214/0.7059  & 0.6227/0.6394
\\ \hline
\end{tabular}
\end{table*}

\subsubsection{The effect of frequency movement.}
\tcr{To validate the effectiveness of our proposed frequency movement, we have a thorough investigation for the three steps in the frequency movement, \ieno, \textit{warm up with pre-training}, \textit{frequency movement-based selection}, and \textit{perturbation-based fine-tuning}, respectively. As shown in Table~\ref{tab:all_ablation}, we can observe that the performance drops without the first step ``warm up with pre-training", since the model is not well-trained and cause the side effect for the frequency selection. For instance, the benefits might be from the better training of the BIQA model instead of the frequency selection. Another crucial step  ``perturbation-based finetuning" achieves an ideal performance improvement. That is because the perturbation provides a tolerance for the frequency selection, which can adjust the frequency component adaptively if there existed better frequency components around the selected frequency band. We also investigate the necessity of our proposed frequency movement in frequency alignment. Concretely, we replace our proposed frequency movement with a simple frequency selection strategy, which exploits the commonly-used Gumbel Softmax~\cite{jang2016categorical} to adaptively select the optimal frequency bands. By comparing the $4^{th}$ line and the $6^{th}$ line, we can find that our proposed frequency movement outperforms the simple Gamble-based frequency selection by 0.0366/0.0206 on SROCC/PLCC. This reveals the effectiveness of our proposed frequency movement. We also conduct the ablation study for the self-supervised loss $\mathcal{L}_T$ in ~\cite{SUFDA}. We can find that this self-supervised loss can bring an excellent average gain of 0.0551/0.0396 on SROCC/PLCC.}

 \subsubsection{The effects of different trajectories and move interval}
In this part, we study three different trajectories for frequency movement in Fig.~\ref{fig:movement_trajectory}, including up-to-down, left-to-right, and Zig-zag Scanning. We show the experimental results in Table~\ref{tab:trajectory}, where we can observe that the left-to-right trajectory achieves the best performance on two cross-domain settings. 
\tct{The different experimental results among the three types of trajectories can be attributed to the transferability distribution of the different frequencies. In principle, the cumulative transferability of our FreqAlign is determined by the transferability of individual frequencies within the sliding window. As shown in Fig.~\ref{fig:priliminary}, frequencies with heightened transferability predominantly align along the lines. By scanning frequencies along the trajectory "left-to-right", the sliding window is able to cover a greater number of high-transferability frequencies, optimizing the overall transfer efficiency.}

We also investigate the effects of the different lengths of one move interval.  We adopt the 6, 8, 10, 12, 14, 16, and 18 epochs as the length of one move interval. \tct{The SROCC curve and PLCC curve} are shown in Fig.~\ref{fig:ablation_for_interval_group}, which reveals that 10 epochs are optimal as one move interval for domain adaptation of large datasets (KADID10k$\rightarrow$KonIQ-10k and KonIQ-10k $\rightarrow$KADID10k), and 14 epochs are optimal for small datasets (CSIQ$\rightarrow$CLIVE and CLIVE$\rightarrow$CSIQ)

\subsubsection{The effects of the number of frequencies in the sliding window} 
\begin{figure}
    \centering
\includegraphics[width=1\linewidth]{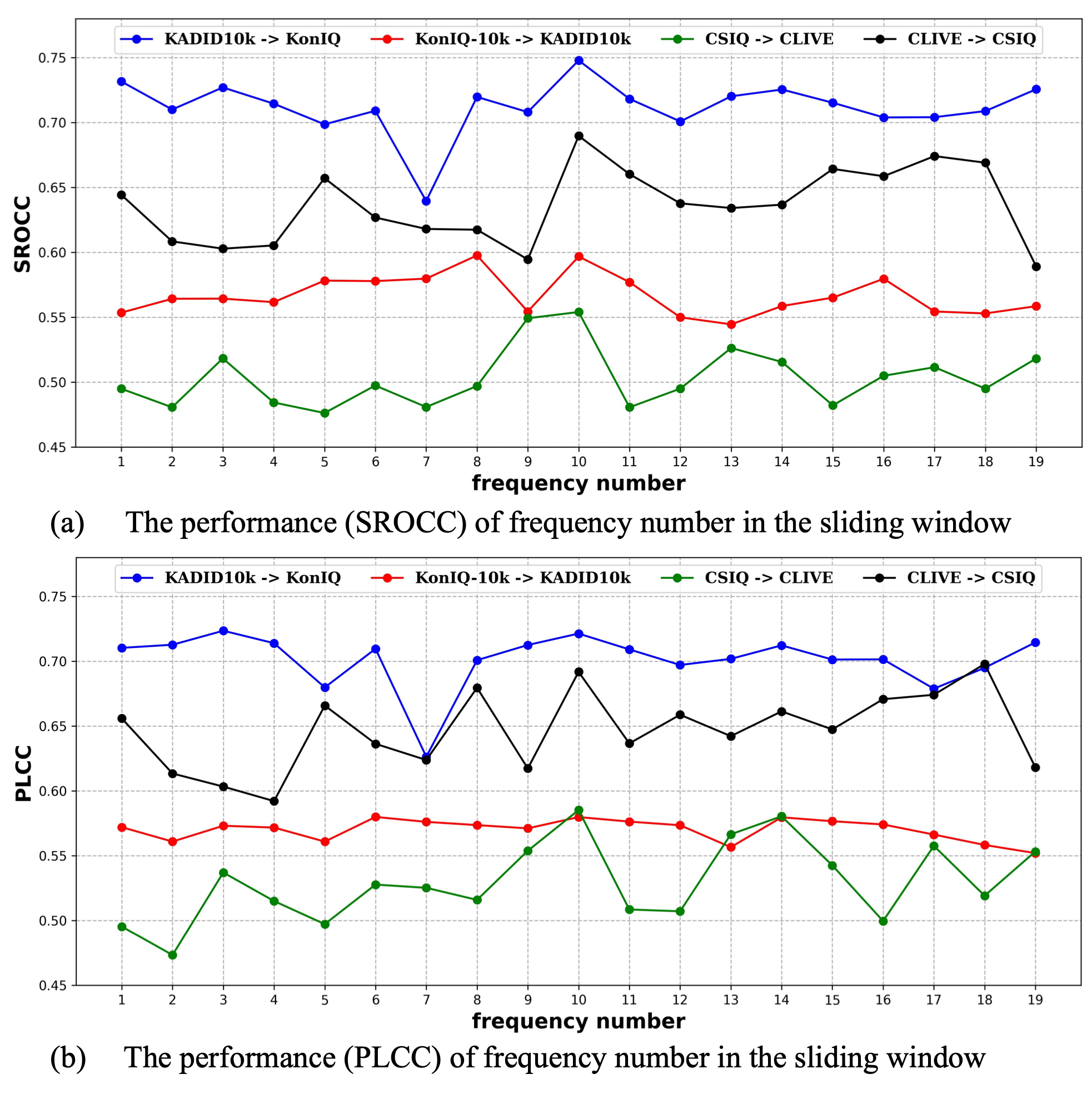}
    \caption{\tct{The performances of different numbers of frequencies in one band (\ieno, in the sliding window).}}
    \label{fig:ablation_for_slt}
\end{figure}

\tcr{To investigate the effects of sliding window size on the frequency movement, we conduct the frequency alignment with different numbers of frequencies, \tct{which ranges from 1 to 19.}
As shown in Fig.~\ref{fig:ablation_for_slt}, we can observe that: with the increase of the number of frequencies, the average performance of four cross-domain scenarios increases first and then drops at the number of 10. The reason for that we believe is:  multiple frequencies contain more useful perception knowledge than few frequencies, but too many frequencies will bring difficulties for the optimization. In this paper, the number of frequencies in one sliding window is set as 10. }

\subsubsection{The effects of transferability metric}
\tcr{As shown in Table~\ref{tab:ab_metric}, we compare three commonly-used metrics for the measurement of transferability, including MMD~\cite{mmd}, CORAL~\cite{CORAL}, and Adversarial loss~\cite{DANN1}. Particularly, MMD is used to measure the mean maximum mean distance between the source and target features. CORAL~\cite{CORAL} measures the transferability with the covariance distance between the features of two domains. 
The experimental results show that these three metrics achieve comparable performance. And the MMD metric is the optimal one, which is used in our frequency alignment.}
\begin{figure}[htp]
    \centering
\includegraphics[width=0.5\textwidth]{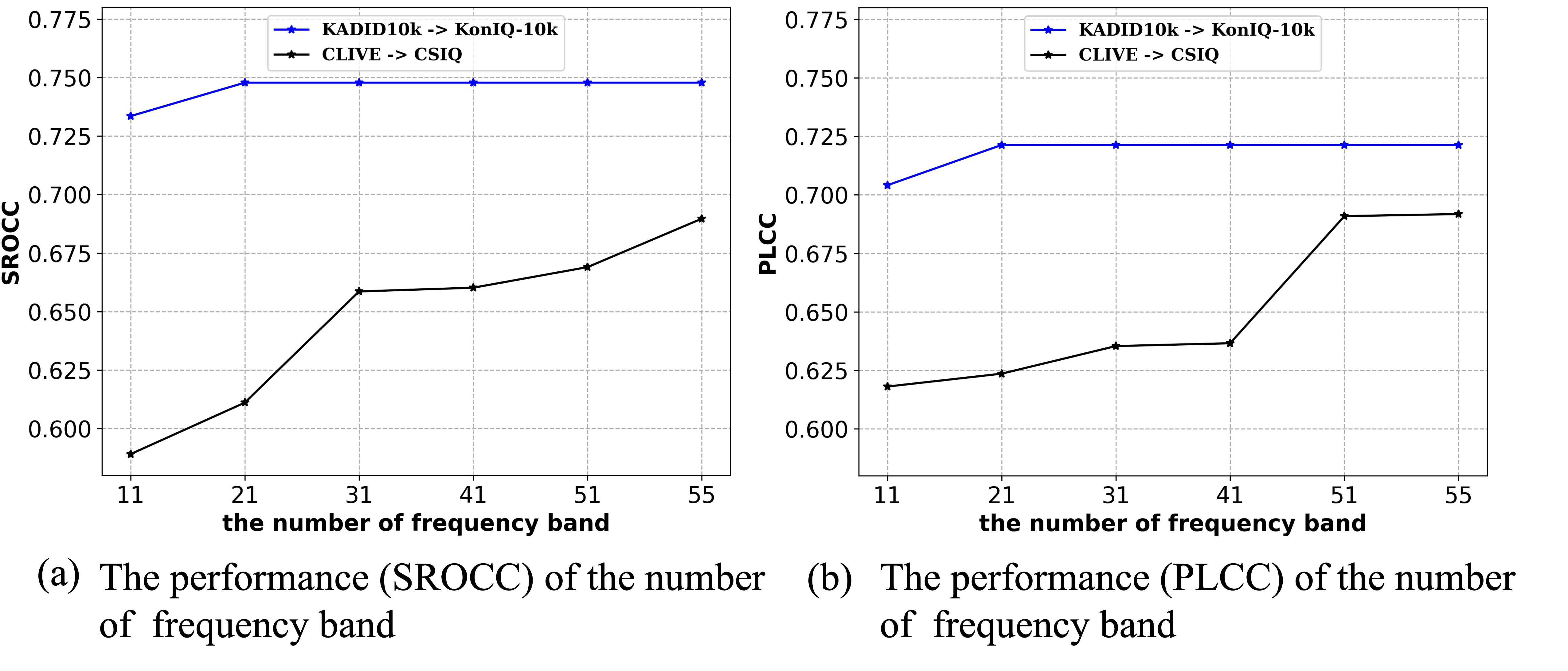}
    \caption{\tct{Ablation studies for the number of frequency bands.}}
    \label{fig:ablation_for_band}
\end{figure}
\subsubsection{\tct{The effects of the number of frequency bands}}
\tct{Intuitively, to select the optimal frequency band, we have to achieve frequency movement-based selection in all frequencies. Therefore, the number of frequency bands $n$ is decided by the frequency number $m$ in one sliding window as $n=64-m+1$. However, for one cross-domain setting, the optimal frequency band might not cover the last frequency in the frequency matrix.
Therefore, we conduct the ablation studies for the number of frequency bands. Concretely, we set the number of frequency bands as 11, 21, 31, 41, 51, and 55 (\ieno, full frequencies), respectively. The experimental results are shown in Fig.~\ref{fig:ablation_for_band}. We can observe that the optimal frequency band can be selected by only traveling the first 21 frequency bands for KADID10k$\xrightarrow[]{}$KonIQ-10k, while CLIVE$\xrightarrow[]{}$CSIQ is required to traverse all frequency bands for selecting optimal frequency band. Notably, in real-world applications, we cannot identify which frequency band is optimal for new cross-domain settings, and it is best to traverse all frequencies for better transferability.}

\subsection{Extension to the UDA of Aesthetic Quality Assessment}
\tct{We also have validated our FreqAlign in 
 aesthetic quality assessment as an extension. We have selected two prominent aesthetic image datasets for our evaluation: AVA~\cite{murray2012ava} and TAD66k~\cite{TAnet}. Specifically, AVA is a widely recognized dataset in aesthetic image research, encompassing over 250,000 images. It employs a voting scale that varies from 1 to 10. On the other hand, TAD66k is a theme-oriented dataset with a collection of 66,000 images across 47 popular themes. The aesthetic scores for TAD66k vary between 1.13 and 9.46.
For evaluation, we also adopted the same metrics used in image quality assessment: SROCC and PLCC. }

\tct{We selected two baseline models, including ResNet-18 and TANet~\cite{TAnet}. Here, TANet extracts multiple priors for aesthetic quality assessment, including semantic prior, high-level color information, and aesthetic details, which is significantly different from ResNet-18. To ensure the fairness of the experiments, we incorporate our FreqAlign to these two baselines, respectively, and compare them with the existing BIQA method MUSIQ~\cite{MUSIQ} and several typical UDA-based methods with ResNet-18, including DANN~\cite{DANN}, RankDA~\cite{RankDA}, and UCDA~\cite{UCDA}. All models are retrained with the datasets on aesthetic quality assessment.} 

\tct{The experimental results are presented in Table~\ref{tab:compareava}. Comparing Baseline (ResNet-18), DANN, and our FreqAlign in the $5^{th}$, $4^{th}$, and $6^{th}$ row of Table~\ref{tab:compareava}, it is obvious that our proposed freqAlign exhibits a superior average gain of 0.0579/0.0544 over the baseline, and outperforms DANN~\cite{DANN} by an average margin of 0.0099/0.0034 in terms of SROCC/PLCC. Moreover,
from Table~\ref{tab:compareava}, we can find that TANet achieves better generalization capability on aesthetic quality assessment than MUSIQ and ResNet-18, revealing the effects of auxiliary priors. By incorporating our FreqAlign, we can also achieve an average gain of 0.0106/0.0057 for the generalization of TANet. This also demonstrates the robustness of our FreqAlign across different backbones. The above experiments reveal that our FreqAlign is also suitable and effective for the UDA of aesthetic quality assessment.}

\begin{table}[htp]
\centering
\caption{\tct{The performance comparison with the state-of-the-art BIQA methods on cross aesthetic image domains. The best performance is bold.}}
\label{tab:compareava}
\setlength{\tabcolsep}{0.5mm}{
\begin{tabular}{c|cc|c}
\hline\hline
              & \begin{tabular}[c]{@{}c@{}}AVA$\rightarrow$TAD66k\\ SROCC/PLCC\end{tabular} & \begin{tabular}[c]{@{}c@{}}TAD66k$\rightarrow$AVA\\   SROCC/PLCC\end{tabular} &Average\\ \hline

MUSIQ~\cite{MUSIQ}          &  0.2158/0.2276                                                         &0.35971/0.3701                                                        & 0.2877/0.2988
                         \\
RankDA~\cite{RankDA}    &0.2790/0.2763     
  &0.3410/0.3791
                           &0.3100/0.3277 \\ 
UCDA~\cite{UCDA}    &0.2965/0.3167     
  &0.4723/0.4807
                           & 0.3844/0.3987\\ 
DANN~\cite{DANN}       & 0.3964/0.4189                                                        &0.5085/0.5194                                                                & 0.4524/0.4691 
                            \\ \hline \hline
Baseline (ResNet-18) ~\cite{ResNet}     & 0.3465/0.3644                                                          & 0.4623/0.47187                                                         & 0.4044/0.4181\\
+ FreqAlign (ours)         & 0.4117/0.4288                                                            &0.5129/ 0.5162
                           &0.4623/0.4725  \\ \hline \hline
                           
Baseline (TANet)~\cite{TAnet}          &0.4156/0.4406                                                             & 0.5820/0.5855 & 0.4988/0.5130\\
+ FreqAlign (Ours)      &\textbf{0.4283}/ \textbf{0.4471}     
              &  \textbf{0.5905}/\textbf{0.5903}                                        
              & \textbf{0.5094}/\textbf{0.5187}                                          
             \\
\hline 
\end{tabular}}
\end{table}

\section{Conclusion}
\tcr{In this paper, we revisit the previous UDA work of BIQA and find that the previous alignment strategy is usually taken at the low-frequency component, which ignores the excavation of the transferable perception knowledge in other frequencies. To improve perception-oriented transferability, we propose one simple and effective method, termed frequency alignment for the UDA of BIQA. Concretely, we perform the frequency decomposition for the aligned features and select the most informative frequency band for the alignment. To achieve stable and efficient frequency selection, we propose frequency movement with the sliding window, which is divided into three steps, including warm-up with pre-training, frequency movement-based selection, and perturbation-based finetuning. Extensive experiments and abundant ablation studies have demonstrated the effectiveness of our proposed frequency alignment on the UDA of the BIQA under three categories of cross-domain settings, including the synthetic-to-authentic setting,  authentic-to-synthetic setting, and cross-distortion types setting.}

\bibliographystyle{IEEEtran}
\bibliography{references}

\end{document}